%% file: main.tex
\documentclass[journal]{IEEEtran}

\ifCLASSINFOpdf

\else

\fi
\usepackage{nccmath,amssymb}

\usepackage[nocompress]{cite}

\usepackage{amsmath}
\usepackage{colortbl}
\usepackage{mathtools}
\usepackage[pdftex]{graphicx}
\usepackage[colorlinks=true,linkcolor=red,breaklinks=True,citecolor=red]{hyperref}
\PassOptionsToPackage{hyphens}{url}

\PassOptionsToPackage{dvipsnames}{xcolor}

\usepackage[font=footnotesize]{caption}

\newcommand*{\vv}[1]{\vec{\mkern0mu#1}}

\newcommand\smallarray[1]{{\small$\vv{#1}$}}
\newcommand\smallset[1]{{\normalsize$\bar{#1}$}}
\newcommand\scriptset[1]{{\scriptsize$\bar{#1}$}}

\newcommand\condeq[2]{{\scriptsize$\bar{#1} = \bar{#2}$}}
\newcommand\condneq[2]{{\scriptsize$\bar{#1} \neq \bar{#2}$}}

\usepackage{adjustbox}
\usepackage[linesnumbered,algo2e,ruled]{algorithm2e}
\usepackage{multirow}
\usepackage{booktabs,subcaption,dcolumn}
\usepackage{tikz}
\usepackage{enumitem}
\usepackage{tabularx}
\usepackage{amsfonts}
\usepackage{pifont}
\usepackage{svg}
\usepackage{soul}
\PassOptionsToPackage{usenames,dvipsnames}{xcolor}

\newcolumntype{?}{!{\vrule width 1.5pt}}

\newcommand\RQ[1]{{\fontfamily{cmss}\selectfont #1}}
\newcommand{\C}[1]{{{\small{\textsf{C#1}}}}}

\newcommand{\captionC}[1]{{{\scriptsize{\textsf{C#1}}}}}

\newcommand\dataset[1]{{\fontfamily{pcr}\selectfont {\small #1}}}

\newcommand\captiondataset[1]{{\fontfamily{pcr}\selectfont {\scriptsize #1}}}

\usepackage{mathtools}
\newtagform{show_eq}{(Exp.\ }{)}
\usetagform{show_eq}

\def\BibTeX{{\rm B\kern-.05em{\sc i\kern-.025em b}\kern-.08em
    T\kern-.1667em\lower.7ex\hbox{E}\kern-.125emX}}
\begin{document}

\title{The Cross-evaluation of Machine Learning-based Network Intrusion Detection Systems}

\author{
\IEEEauthorblockN{Giovanni Apruzzese\IEEEauthorrefmark{1}, Luca Pajola\IEEEauthorrefmark{2}, Mauro Conti\IEEEauthorrefmark{2}\IEEEauthorrefmark{3}\\
giovanni.apruzzese@uni.li, \{pajola, conti\}@math.unipd.it}\\
\IEEEauthorblockA{\IEEEauthorrefmark{1}\textit{Institute of Information Systems -- University of Liechtenstein}}\\
\IEEEauthorblockA{\IEEEauthorrefmark{2}\textit{Department of Mathematics -- University of Padua, Italy}}\\
\IEEEauthorrefmark{3}\textit{Faculty of Electrical Engineering, Mathematics and Computer Science -- TU Delft}
}

\markboth{IEEE Transactions on Network and Service Management}%
{Shell \MakeLowercase{\textit{et al.}}: Bare Demo of IEEEtran.cls for IEEE Journals}

\maketitle

\input{sections/0-abstract.tex}


\IEEEpeerreviewmaketitle

\input{sections/1-introduction}

\input{sections/2-background}

\input{sections/3-idea}

\input{sections/4-framework}

\input{sections/5-application}

\input{sections/6-demonstration}
\input{sections/7-discussion}

\input{sections/8-conclusions}

\ifCLASSOPTIONcaptionsoff
  \newpage
\fi
\input{bibliography}


\input{biographies/bio}

\appendices
\input{sections/appendix}

\end{document}

%% file: sections/0-abstract.tex
\begin{abstract}
Enhancing Network Intrusion Detection Systems (NIDS) with supervised Machine Learning (ML) is tough. ML-NIDS must be trained and evaluated, operations requiring data where benign and malicious samples are clearly labelled. Such labels demand costly expert knowledge, resulting in a lack of real deployments, as well as on papers always relying on the same outdated data. The situation improved recently, as some efforts disclosed their labelled datasets. However, most past works used such datasets just as a `yet another' testbed, overlooking the added potential provided by such availability.

In contrast, we promote using such existing labelled data to \textit{cross-evaluate} ML-NIDS. Such approach received only limited attention and, due to its complexity, requires a dedicated treatment. We hence propose the first cross-evaluation model. Our model highlights the broader range of realistic use-cases that can be assessed via cross-evaluations, allowing the discovery of still unknown qualities of state-of-the-art ML-NIDS. For instance, their detection surface can be extended---at no additional labelling cost.
However, conducting such cross-evaluations is challenging. Hence, we propose the first framework, XeNIDS, for reliable cross-evaluations based on Network Flows. By using XeNIDS on six well-known datasets, we demonstrate the concealed potential, but also the risks, of cross-evaluations of ML-NIDS. 
\end{abstract}

%% file: sections/1-introduction.tex
\section{Introduction}
\label{sec:introduction}

\IEEEPARstart{M}{achine} Learning (ML) is advancing at a rapid pace (e.g.,~\cite{sidey2019machine,wu2019machine}), and the cybersecurity domain is also looking at ML with great interest~\cite{bresniker2019grand}. 
ML methods can automatically learn to make decisions by using existing data, representing a valuable asset to monitor the increasingly mutating IT environments.

Although ML is already deployed to counter some threats (e.g., malware or phishing~\cite{fleshman2018static, d2020malware, liang2016cracking}), ML methods are still at an early stage for Network Intrusion Detection (NID).
In particular, some Network Intrusion Detection Systems (NIDS) integrate commercial products that use \textit{unsupervised} ML (e.g.,~\cite{Darktrace:CyberML,Lastline:AI}).
Such solutions can be useful to perform correlation analyses or to `detect anomalies', which are ancillary to true intrusion detection tasks (an anomaly is not necessarily an intrusion). 
The full potential of ML can be appreciated only via \textit{supervised} methods, which assume the existence of \textit{labels} that associate each sample to its ground truth~\cite{Sommer:Outside}. Specifically in NID, by creating a training dataset where the samples are distinguished between \textit{benign} and \textit{malicious}, it is possible to develop a fully autonomous Machine Learning-based Network Intrusion Detection System (ML-NIDS). 

Deployment of ML-NIDS involves two stages: the system must first be developed (i.e., \textit{trained}), and it must then be \textit{evaluated}, because any security system that has not been tested is dangerous~\cite{biggio2013security}. Both of these stages require \textit{large} amounts of \textit{labelled} data, which can only be collected via the supervision of a human that associates (and verifies) each sample to its ground truth~\cite{Apruzzese:Deep}. While such verifications are simple in some applications (e.g., any layman can distinguish a legitimate from a phishing website), the inspection of network data requires expert knowledge--which is expensive~\cite{miller2016reviewer}. To aggravate the problem, a network can be targeted by many attacks, \textit{each} of which must be labelled to assess the detection capabilities of a ML-NIDS.
As a result, the inevitable and costly necessity of comprehensive labelled datasets (usually numbering millions of samples~\cite{ring2019survey}) discourages deployment of ML-NIDS. We note that this problem also extends to \textit{research}. For more than a decade, the only publicly available dataset for ML-NIDS was the KDD99, leading to a plethora of works always trained and evaluated on such dataset---usually with perfect performance (e.g.,~\cite{mishra2018detailed}).

To address the lack of labelled data, recent researches on ML-NIDS openly released their datasets (e.g.,~\cite{ring2019survey, CICIDS2018:Dataset, Garcia:CTU}), an effort appreciated by related literature (e.g.,~\cite{mishra2018detailed, apruzzese2020deep, injadat2020multi}). However, most prior works used such datasets as an additional testbed for their proposals. As a result, such works only confirmed what was already known: that by training a ML-NIDS on a (large) dataset, such ML-NIDS will detect the attacks contained in such dataset. This is because the primary objective was to `outperform' the state-of-the-art, resulting in incremental contributions that do not foster realistic deployments.
In this paper, we aim to broaden such limited scope.

Inspired by a recent paper by Pontes et al.~\cite{pontes2021potts}, we observe that the current availability of labelled datasets could be better exploited by ML-NIDS researches. Specifically, we endorse the idea of \textit{cross-evaluating} ML-NIDS by using malicious samples captured in different network datasets.\footnote{We stress that our term `cross-evaluation' denotes a different concept than the term `cross-validation' commonly used in ML researches~\cite{schaffer1993selecting}.} By performing such cross-evaluations, it is possible to gauge additional properties of ML-NIDS, allowing a better understanding of the state-of-the-art at no extra labelling cost.

To the best of our knowledge, this is the first effort that focuses on the opportunity provided by cross-evaluations of ML-NIDS. As such, our primary goal is the definition of a data-agnostic \textit{model} that allows to represent such cross-evaluations. Indeed, using samples from different networks is not straightforward: as stated by Sommer and Paxson~\cite{Sommer:Outside}, each network has ``immense variability'', suggesting that cross-evaluations have intrinsic risks that must be known to avoid deployments of unreliable ML-NIDS. Our model acknowledges such risks, but also highlights the \textit{benefits} that can be brought by cross-evaluations of ML-NIDS. Such benefits come in the form of additional types of `contexts' that can be reproduced in research environments, each representing a distinct realistic use-case. Specifically, our model highlights the limited scope of the state-of-the-art, whose fixed evaluation methodology can only cover 2 contexts, whereas cross-evaluations can span over up to 10 different contexts. Such broad range evidences the concealed potential of the core idea at the base of our paper. 

As stated by Biggio et al.~\cite{biggio2013security}, ML systems for cybersecurity must be assessed \textit{in advance}. Therefore, \textit{proactive} cross-evaluations must take into account all the pitfalls highlighted by our model. To further promote our proposal, we develop the first framework for cross-evaluations of ML-NIDS, XeNIDS. Our framework aims to overcome the intrinsic challenges of cross-evaluations, while allowing the reproduction of all contexts enabled by our model. Specifically, XeNIDS focuses on NetFlow data, which is popular in the ML-NIDS community (e.g.,~\cite{sarhan2020netflow, pontes2021potts, ring2019survey}). However, using NetFlows from different environments is tough: such data can be generated in many ways, resulting in heterogeneous formats that may lead to unreliable ML-NIDS. We address this issue via an \textit{original} interpretation of NetFlows w.r.t. ML. Using this interpretation, we provide the guidelines that can increase the reliability of the results provided by XeNIDS.

As an instructive \textit{demonstration}, we use XeNIDS to perform a large cross-evaluation of ML-NIDS spanning over 6 well-known and recent datasets. We aim to reproduce realistic use cases, which can be assessed via three different context types enabled by our model. Specifically, we first consider the `baseline' context commonly adopted by prior work, and show that XeNIDS yields the same performance as the state-of-the-art. Then, we assess the context where a ML-NIDS is \textit{tested} on malicious samples originating from different networks; such use-case was also investigated in~\cite{pontes2021potts}, and XeNIDS matches their performance. Finally, we assess the context where the ML-NIDS is \textit{trained and tested} on malicious samples from different networks, showing a dramatic performance increase. As a final contribution of this paper, we provide an in-depth \textit{analysis} of these results, where we investigate their reliability for practical deployments.

\textit{Contribution and Organization.}
This is the first paper that addresses the problem of cross-evaluations of ML-NIDS. As such, the specific contributions are as follows.
\begin{itemize}
    \item We present the first data-agnostic model that conceptualizes the problem of cross-evaluation of ML-NIDS.
    \item We use our model to showcase the benefits and challenges of such cross-evaluations. 
    \item We propose XeNIDS, the first framework for reliable cross-evaluations focused on NetFlow data.
    \item We demonstrate all of the above by cross-evaluating ML-NIDS over 6 distinct well-known datasets, and analyzing the results' reliability.
\end{itemize}
The remainder of the paper has the following structure.
We motivate our paper in §\ref{sec:background}. 
We define our cross-evaluation model in §\ref{sec:idea}.
We describe our XeNIDS framework in §\ref{sec:framework}.
We explain the application of XeNIDS in §\ref{sec:application}.
We present our demonstration in §\ref{sec:demonstration}.
We discuss the results in §\ref{sec:discussion}.
We conclude our paper in §\ref{sec:7-conclusions}.

%% file: sections/2-background.tex
\section{Background}
\label{sec:background}
This work lies at the intersection of Machine Learning and Network Intrusion Detection. 
We first provide some preliminary information on these two areas (§\ref{ssec:mlnids}). Then, we explain the motivation (§\ref{ssec:motivation}) of our paper. Finally, we compare this effort with related work (§\ref{ssec:related}).

\subsection{Network Intrusion Detection and Machine Learning}
\label{ssec:mlnids}
The so-called security lifecycle spans over three activities: prevention, detection, reaction~\cite{williams2018engineering}.
However, the prevention of any cyber-attack is an impossible task, while the reaction phase assumes that most of the damage has already taken place. For this reason, proposals focusing on the detection step have received much more attention, as timely and accurate identifications of cyber threats can significantly mitigate the effects of an offensive campaign~\cite{Pierazzi:Online}.

\begin{figure}[!htbp]
    \centering
    \includegraphics[width=0.95\columnwidth]{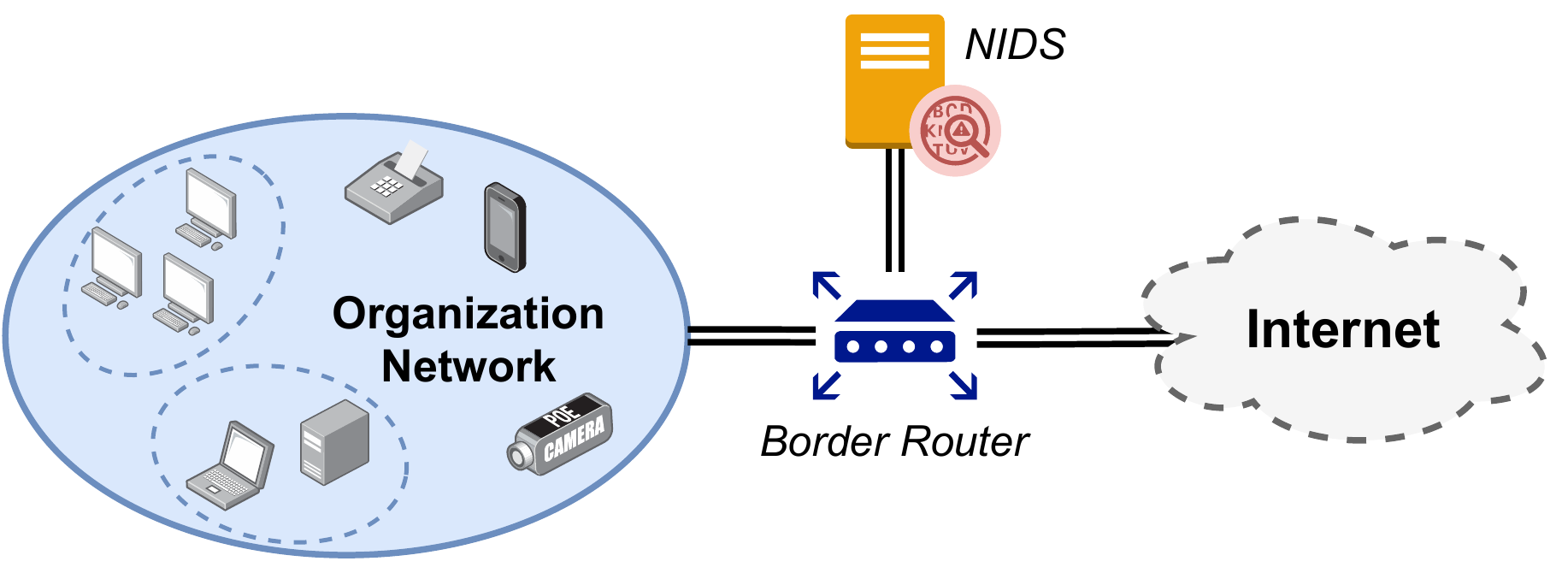}
    \caption{Typical NIDS scenario. The network of the organization can be composed of multiple subnetworks. The outgoing traffic passes through a border router which forwards such traffic to the internet, but also to a NIDS. The NIDS analyzes the traffic and, if necessary, raises some alerts.}
    \label{fig:nids}
\end{figure}

In the specific domain of network security (which is of interest to this paper), the detection of such malicious events is devoted to Network Intrusion Detection Systems (NIDS). We provide a schematic representation of the typical NIDS deployment in Fig.~\ref{fig:nids}, where a NIDS inspects the traffic generated by the monitored network (and all of its subnetworks). A NIDS can leverage two distinct detection paradigms, which are based either on \textit{fixed rules} or on \textit{data-driven} methods~\cite{Buczak:Survey}. The former requires human operators that write specific rules (or signatures) that denote a specific threat, and exhibit high performance against \textit{known} and \textit{static} threats whose behavior is captured by the hardcoded rules. On the other hand, the latter leverage automatic data analyses and can detect even \textit{unknown} and \textit{mutating} threats if they present similarities with previously known samples--potentially at the expense of higher false-positive rates. 

The increased growth of data alongside improvements in collaborative computing resulted in a huge interest in data-driven NIDS, specifically employing machine learning methods~\cite{Apruzzese:Deep,bresniker2019grand}. Such methods involve a \textit{training} phase where the ML model learns to make decisions from existing data. 
However, without some reference information, it is not possible to control what the ML model is actually `learning'~\cite{joyce2021framework}. To specifically address detection (i.e., classification) problems, the training data must be separated into \textit{benign} and \textit{malicious} samples. In such circumstances, it is possible to develop autonomous ML-NIDS exploiting supervised ML methods. Such `supervision' comes in the form of a human that must associate each sample in the training data to its ground truth, i.e., a \textit{label}~\cite{Sommer:Outside}.

In some domains, labelling is simple (e.g., the popular captchas~\cite{bursztein2011text}) or labelled data can be used for a long period of time (e.g., ImageNet was collected in 2009 and is still widely used today~\cite{you2018imagenet}). However, the Cybersecurity domain is different: according to Miller et al.~\cite{miller2016reviewer}, a company can only label 80 malware samples per day. Specifically in NID, ground truth verification of network data is complex~\cite{min2018ids}, and the concept drift problem requires any ML-NIDS to be continuously updated with new--labelled--data~\cite{jordaney2017transcend}. 
To aggravate this problem, deployment of any security system requires \textit{proactive} evaluations conducted \textit{in advance}, to avoid introducing a weak link in the security chain~\cite{biggio2013security}. Hence, in the case of ML-NIDS, labelled data must be obtained both for the initial training, as well as for such evaluation.

The scarcity of labelled data for NID affected both research and practice~\cite{ahmad2021network}, with an overall lack of ML-NIDS deployments, as well as a plethora of papers always based on the only publicly available dataset---the KDD99~\cite{mishra2018detailed}.

\subsection{Motivation: Mixing Network Data}
\label{ssec:motivation}
The successes of ML renewed the interest of the NID community in these methods, and in recent years, many labelled datasets were made openly accessible (a survey is in~\cite{ring2019survey}). However, most related work simply used such data as an `additional' setting to perform their experiments. In contrast, in this paper we promote a different approach, based on mixing different network data to cross-evaluate ML-NIDS. 
Such opportunity, fostered by the recent availability of NID datasets, is of interest both for research and practice. 
Let us explain how mixing network data can assist ML-NIDS deployment. We first by present some high level applications (§\ref{sssec:highlevel}), and then provide a more specific use-case (§\ref{sssec:usecase}).

\subsubsection{Applications and advantages}
\label{sssec:highlevel}
Mixing data from different networks is useful to augment pre-existing datasets that contain an insufficient amount of labelled samples to develop ML-NIDS. It is also useful to assess the \RQ{generalization} capabilities of a ML-NIDS against `novel' attacks not included in the training set (as very recently done by~\cite{pontes2021potts}). Such `novel' samples can also be used \RQ{extend} the detection surface of the ML-NIDS by injecting them in the training set of the ML-NIDS. Similar strategies are particularly relevant to protect against the so-called `adversarial attacks' which can evade traditional ML-NIDS~\cite{doriguzzi2020lucid}: the (new) training data can be leveraged for \textit{adversarial training}, therefore realizing robust ML systems that can detect even subtle perturbations~\cite{abou2020evaluation}. In this context, mixing diverse datasets facilitates the application of \textit{ensemble} techniques (e.g.,~\cite{zhang2020tiki}), further increasing the resilience of ML-NIDS.

As stated by Biggio and Roli, empirical evaluations are always necessary for real deployments~\cite{Biggio:Wild}. In this context, cross-evaluations are advantageous due to their low opportunity cost---especially when using publicly available data. Indeed, we observe that great attention has been given to \textit{data sharing} platforms (e.g.,~\cite{horak2019gdpr}), and cross-evaluations could greatly benefit from dedicated `banks' of NID data (e.g.,~\cite{spagnolettia2020digital}): it is true that the cybersecurity domain has high confidentiality, but anonymization techniques exists~\cite{ramaswamy2007high}, and some recent solutions in \textit{federated learning} overcame privacy issues (e.g.~\cite{dayan2021federated}). Finally, cross-evaluations can involve even \textit{unsupervised} ML methods (e.g., anomaly detectors~\cite{falcao2019quantitative}), which represent the majority of currently deployed ML techniques for NIDS. Although unsupervised methods would not benefit from the `cheap' labelling, they can still take advantage of the data diversity of different networks to assess (or improve) their generalization capabilities.

\subsubsection{Exemplary use-case}
\label{sssec:usecase}
Suppose an organization, $\mathcal{O}$, wants to protect their network, $o$, with a (supervised) ML-NIDS. Hence, $\mathcal{O}$ collects and verifies some \textit{benign} traffic data, $N$, from their network $o$. However, ML-NIDS also require \textit{malicious} data, $M$.
The following can happen w.r.t. such $M$:
\begin{itemize}
    \item $\mathcal{O}$ may not have any $M$ generated in their network $o$. Hence, $\mathcal{O}$ can `use' some $M$ generated in a different network than $o$ -- potentially of another organization. 
    \item $\mathcal{O}$ may have some $M$ generated in $o$, obtained, e.g., by monitoring the behaviour of `known' infected machines.
\end{itemize}
Therefore, $\mathcal{O}$ can use such $N$ and $M$ to develop any ML model which, if it obtains appreciable performance, will be integrated in their security system as a ML-NIDS that can detect the attacks in $M$.
Having an operational ML-NIDS, $\mathcal{O}$ may be willing to assess whether such system can detect attacks not included in their $M$, which can potentially target the network $o$ monitored by the ML-NIDS. To this end, $\mathcal{O}$ can use a \textit{small} set of malicious data originating from a network different than $o$, and containing different attacks than the ones `learned' by their ML-NIDS. By using such malicious data to \textit{evaluate} the ML-NIDS, $\mathcal{O}$ can assess the \RQ{generalization} capabilities of their solution. If the assessment shows a weakness of the ML-NIDS, $\mathcal{O}$ may acquire a \textit{larger} set of such malicious data to \RQ{extend} the detection capabilities of their ML-NIDS, by using such data in the \textit{training} stage.
We will use the abovementioned example as basis for our demonstration in §\ref{sec:demonstration}.

\subsection{Related Work}
\label{ssec:related}
The idea of cross-evaluating ML-NIDS on different datasets is not new. For instance, the authors of~\cite{catillo2021usb} propose a novel IDS dataset that can be used to evaluate the `transferability' of ML-NIDS, but they do not provide any detailed analysis nor original experiment. Similarly, Pontes et al.~\cite{pontes2021potts} use a ML-NIDS trained on \dataset{IDS18} against \dataset{DDoS19}. However,~\cite{pontes2021potts} simply limit to \textit{test} a novel method on a different dataset, and do not analyze the problem of `cross-evaluations' as a whole, hence not allowing to highlight the benefits and limitations of such opportunity. For instance, cross-evaluations can also involve modifications of the \textit{training} data, which is not covered by~\cite{pontes2021potts} and which is a case included in our demonstration. 

Most prior works on ML-NIDS only assess their proposals in a single `context', that is, the training and evaluation use the same dataset. For instance, the authors of~\cite{bansal2017comparative} propose botnet detectors trained and tested on \dataset{CTU13}. In~\cite{vinayakumar2019deep} a ML-NIDS focusing on different attacks is assessed on the \dataset{IDS18} dataset. To give a practical explanation, such methodology only allows determining that ``the approach in~\cite{vinayakumar2019deep} is effective on the network captured by the \dataset{IDS18} dataset, against the attacks contained in the \dataset{IDS18} dataset''. Other works may consider more datasets (e.g.,~\cite{divekar2018benchmarking, injadat2020multi, magan2020towards, sarhan2020netflow}), but the problem remains because the assessments are carried out independently on each dataset.
Furthermore, all these works highlight that ML-NIDS require large (labelled) datasets--further motivating the need to explore novel solutions that mitigate the lack of labelled data. Among these, we mention \textit{semisupervised} ML approaches (e.g.~\cite{zhang2021network, min2018ids}), which combine unlabelled with labelled data, and are hence orthogonal to our work. 

A closely related research effort is~\cite{cordero2021generating}, proposing a low-level software toolkit for analyzing NID datasets, forcing the user to abide to its constrained logic. For instance, it only works with data in the form of packet captures (PCAP), which require huge amounts of storage space and whose payload is often encrypted, making such data impractical to share (and, also, to analyze). In contrast, our proposed model is agnostic of the source data format (as long as there is some compatibility); moreover, our proposed framework operates on Network Flows (NetFlows), which represent a higher level than PCAP, making it flexible and extendible also to PCAP data---while not sacrificing performance~\cite{Bilge:Disclosure, bansal2017comparative ,apruzzese2020deep}. 

We conclude that the idea of cross-evaluating ML-NIDS received only limited attention so far, and its opportunities and risks are still unknown. This is because no past research truly addressed such a problem---representing the core of this paper. Our intention is to provide a complete understanding of all the pros and cons related to cross-evaluations of ML-NIDS.

%% file: sections/3-idea.tex
\section{Modelling the Cross-evaluation of ML-NIDS}
\label{sec:idea}
The intuition at the base of our work is to leverage \textit{existing} NID datasets, with the goal of cross-evaluating ML-NIDS using samples from mixed networks.
Such idea is grounded on the following observation (also implicitly adopted by~\cite{pontes2021potts}), which extends the takeaways by Sommer et Paxson~\cite{Sommer:Outside}: although every network is unique, the malicious behavior of network attacks is independent of the target network. For instance, Denial of Service (DoS) attacks always involve either a large amount of communications with minimal entity, or a smaller set of communications but with a larger entity -- both happening in a short time frame~\cite{hou2018machine}. Similarly, a machine infected by Botnet malware will periodically contact the CnC server, irrespective of what is happening in the `compromised' network~\cite{Apruzzese:Periodic}. Hence, such malicious behaviours can be used by a ML-NIDS to distinguish benign from malicious activities, regardless of the target network.

As the first effort to investigate this opportunity, we must design a \textit{model} (§\ref{ssec:principles}) that allows to highlight its \textit{benefits} (§\ref{ssec:benefits}) as well as its intrinsic \textit{challenges and risks} (§\ref{ssec:challenges}).

\subsection{Proposed Model Design}
\label{ssec:principles}
We now introduce all the prerequisites to describe our cross-evaluation model.

Let $\mathbb{D}$ be a set of NID datasets which we denote as follows:
\begin{align*}
    \mathbb{D}=(D_1, D_2, ..., D_n), n\geq2,    
\end{align*}
\noindent
where $n$ represents the cardinality of $\mathbb{D}$, and $D_i$ represents a dataset collected in a given network $i$.
Without loss of generality, we assume that each $D_i \in \mathbb{D}$ originates from a unique network environment---potentially, $\mathbb{D}$ can include datasets representing distinct sub-networks within a larger network. %
Hence, in the remainder we use $D_i$ and $D_j$~($i\neq j$) to denote two datasets of $\mathbb{D}$ originating from two distinct networks (i.e., $i$ and $j$).
The \textit{information} captured by each dataset in $\mathbb{D}$ must allow one to use any subset of $\mathbb{D}$ and derive a \textit{set of common features} from such subset. \footnote{For instance, it is possible that $D_i$ comes as PCAP traces, and $D_j$ comes as NetFlows: in this case, the PCAP of $D_i$ can be processed to derive the NetFlows features of $D_j$. Similarly, two datasets $D_i$ and $D_j$ can contain NetFlows generated with different software: in this case, the features shared by $D_i$ and $D_j$ can represent the common set.}

Because our focus is on supervised ML for detection problems, each $D_i \in \mathbb{D}$ must be provided with \textit{ground truth} distinguishing benign from malicious data.
Hence, each dataset $D_i$ can be seen as a composition of $N_i$, denoting the \textit{benign} data of network $i$, and $M_i$, denoting the \textit{malicious} data of network $i$. A dataset $D_i$ can contain only malicious (or only benign) data; however, across all $\mathbb{D}$ there must be at least a pair $D_i$, $D_j$ for which $N_i \neq \varnothing$ and $M_j \neq \varnothing$.
We denote with $\mu$ the number of \textit{all} malicious classes contained in the entire $\mathbb{D}$. This is because any $D_i$ can have a $M_i$ with a variable number of attacks (i.e., Botnet, DoS, etc), which may overlap (or not) with those in a different $M_j$. Therefore, every $M_i$ can be seen as an array of $\mu$ elements $M_i\!=\!(M_i^1, M_i^2, ..., M_i^\mu)$, some of which can be empty if another dataset $D_j$ has malicious classes not contained in $D_i$.
Let $\mathbb{N}$ denote the set of all benign samples, and $\mathbb{M}$ denote the set of all malicious samples. 

\begin{figure}[!htbp]
    \centering
    \includegraphics[width=0.7\columnwidth]{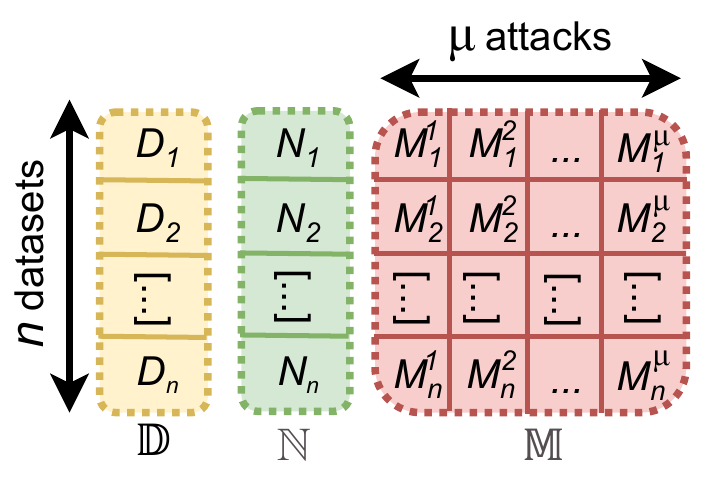}
    \caption{Proposed cross-evaluation model, representing $\mathbb{D}$, $\mathbb{N}$ and $\mathbb{M}$. \textbf{Example}: assume that $\mathbb{D}$ has three datasets: $D_1, D_2, D_3$ (hence $n$=3). Assume that: $D_1$ has some benign samples but no malicious samples; $D_2$ has no benign samples, but malicious samples of a \textit{botnet} attack, and also some malicious samples of a \textit{portscan} attack; $D_3$ has some benign samples, alongside some malicious samples of the same \textit{botnet} attack as $D_2$, but also some malicious samples of a \textit{DoS} attack. In this case, $\mathbb{N}$ will contain three elements: $N_1$, $N_2$, $N_3$, with $N_2$ being empty. On the other hand, there are three malicious classes ($\mu$=3) hence $\mathbb{M}$ will have nine elements: $M_2^1$ and $M_3^1$ (the botnet attack shared by $D_2$ and $D_3$) as well as $M_2^2$ (the portscan attack in $D_2$) and $M_3^3$ (the DoS attack in $D_3$) will contain some samples; whereas the remaining will be empty (i.e., all those with $M_1$, as well as $M_2^3$ and $M_3^2$).}
    \label{fig:DNM}
\end{figure}

We can visualize our model with the schematic in Fig.~\ref{fig:DNM}, which shows the relationship between $\mathbb{D}$, $\mathbb{N}$ and $\mathbb{M}$.

From Fig.~\ref{fig:DNM}, we observe that all sets have $n$ rows, each denoting a distinct source network dataset. However, while $\mathbb{N}$ has only one column because all benign samples are treated equally, $\mathbb{M}$ has $\mu$ columns representing all the attacks contained in $\mathbb{D}$. We provide an example in the caption of Fig.~\ref{fig:DNM}.

Such design makes our model suitable for 1+$\mu$ classification ML problems, where a sample is either benign, or belongs to one among $\mu$ malicious classes. This automatically covers binary classification ML problems if all malicious classes are treated as a single malicious class (hence, $\mu$=1).

Let us now use our proposed model to explain the benefits brought by cross-evaluations of ML-NIDS.

\subsection{Benefits: additional Contexts}
\label{ssec:benefits}
Deployment of ML components requires a \textit{training} set $T$, used to develop a ML model, and an \textit{evaluation} set $E$, used to assess the performance of such model. Hence, our idea is composing $T$ and $E$ by drawing from $\mathbb{N}$ and $\mathbb{M}$: depending on the draw, a specific `context' is created that can be used to cross-evaluate a ML-NIDS. The main benefits provided cross-evaluations of ML-NIDS are due to the increased types of contexts that can be assessed, which can be highlighted with our proposed model.

Because our model is rooted on Sommer and Paxson statement (§\ref{sec:idea}), it is crucial that both $T$ and $E$ use benign samples from the same network\footnote{This also serves to reduce false alarms \textit{after} potential deployments, because the benign samples always have the same source.}, which should represent the environment where the ML-NIDS is to be deployed; hence, let $o$ (standing for `origin') denote such network, and $N_o$ be the corresponding benign samples. Then, we observe that there are many ways to compose $T$ and $E$ by choosing the malicious element from $\mathbb{M}$. Such variability can be modeled through the `matrix' $\mathbb{M}$ in Fig.~\ref{fig:DNM}, by pinpointing which rows and columns are included in $T$ and $E$.
The following can occur:
\begin{itemize}
    \item $T$ (or $E$) can contain malicious samples either from the same or different $o$ (i.e., same or different row than $N_o$);
    \item the malicious samples in $T$ and $E$ can come either from the same or different networks (regardless of $N_o$);
    \item the malicious class(es) in $T$ can be either the same or different than those in $E$ (i.e., same or different columns).
\end{itemize}
In particular, let $t$ and $e$ denote two rows of $\mathbb{M}$; and let $\tau$ and $\varepsilon$ denote two columns of $\mathbb{M}$; we use such notation to identify two elements of $\mathbb{M}$, i.e., $M_t^\tau$ and $M_e^\varepsilon$.

Let \smallarray{t}, \smallarray{e}, \smallarray{\tau}, \smallarray{\varepsilon} be four \textit{ordered arrays}\footnote{Such arrays can be of variable length, but $|\vv{t}|$=$|\vv{\tau}|$ and $|\vv{e}|$=$|\vv{\varepsilon}|$ must be true.}, each denoting multiple columns or rows (e.g., \smallarray{t} contains multiple $t$, i.e., rows) of $\mathbb{M}$.
Let \smallarray{o} be an unary array including only $o$ (representing the benign `origin' network from $\mathbb{N}$).

By following such notation, we can represent the training and evaluation sets, $T$ and $E$, as the functions in Expression~\ref{exp:TE}:
\begin{align}
  \begin{aligned}
    T(\vv{o}, \vv{t}, \vv{\tau}), & & & E(\vv{o}, \vv{e}, \vv{\varepsilon}).
  \end{aligned}
  \label{exp:TE}
\end{align} 
Simply put, $T$ and $E$ are denoted by a single row of $\mathbb{N}$ (i.e., \smallarray{o}), and the rows and columns of all the elements of $\mathbb{M}$ that they include (i.e., \smallarray{t} and \smallarray{\tau} for $T$, while \smallarray{e} and \smallarray{\varepsilon} for $E$).

We can see a \textit{context} as a function of $T$ and $E$. Specifically, a context \C{} is denoted as the following tuple:
\begin{align}
  \begin{aligned}
    \C(T, E) \Rightarrow \C(\vv{o}, \vv{t}, \vv{e}, \vv{\tau}, \vv{\varepsilon}).
  \end{aligned}
  \label{exp:context}
\end{align} 
Depending on the elements from $\mathbb{N}$ and $\mathbb{M}$ included in $T$ and $E$, many contexts can be reproduced, which can be of different \textit{type}. In particular, let \smallset{o}, \smallset{h}, \smallset{e}, \smallset{\tau}, \smallset{\varepsilon} denote the \textit{sets} of the corresponding arrays (each element of a given set is unique). By cross-evaluating ML-NIDS, it is possible to assess 10 different context types, which are denoted by the relationships between \smallset{o}, \smallset{h}, \smallset{e}, \smallset{\tau}, \smallset{\varepsilon}. 

We provide the full list of such context types in Table~\ref{tab:contexts};  we also include a practical example in the caption of Table~\ref{tab:contexts}.
Specifically, for each context type (denoted with a number after the letter \C{}), we report the four conditions denoting the relationships among all the involved components; on the same line of each condition, we describe the consequences on $T$ and $E$; we also provide a concrete use case that explains the application of such type of context. 
We note that all cases where two sets are not equal can be further split in two: when one set is a \textit{superset} of the other; and when they are \textit{disjointed}. 

\input{sections/tab_contexts}

Past works (e.g.,~\cite{apruzzese2020deep, Stevanovic:Botnet}) only considered cases where the `row' was fixed, i.e., where \smallset{o}=\smallset{h}=\smallset{k}, corresponding to the contexts of type \C{1} and \C{2}. Pontes et al.~\cite{pontes2021potts} investigated \C{4}.
In contrast, it is evident from Table~\ref{tab:contexts} that our cross-evaluation model enables the assessment of 7 additional context types, allowing to discern additional qualities of ML-NIDS and corresponding NID datasets. For instance, all the scenarios envisioned in our motivational example (cf. §\ref{ssec:motivation}) can be represented by the context types listed in Table~\ref{tab:contexts}. 

In our demonstration (§\ref{sec:demonstration}), we first assess \C{1} as `baseline' comparison with the state-of-the-art; and then we consider \C{4} (as done by~\cite{pontes2021potts}), and \C{7} (the latter both in its `disjoined' and `extended' variants).

\subsection{Challenges and Risks}
\label{ssec:challenges}
The cross-evaluation of ML-NIDS has high potential, but a superficial application can lead to dangerous consequences---spanning from underwhelming performance to additional security risks. Indeed, mixing data from different networks presents several fundamental issues, which must be known when real ML-NIDS deployments are considered.
We stress that our paper lies at the intersection of diverse research fields (i.e., network traffic analysis, machine learning, cybersecurity) and some of the following issues may be well-known within each field. Considered the scope of our paper, it is meaningful to make the entire community aware of such issues.

We identify the following three performance-related implementation \textit{challenges}:
\begin{enumerate}
    \item \textit{Removing Network Artifacts.} Depending on the considered set of features, some samples may contain  `artifacts' that are unrelated to their benign/malicious nature\footnote{The most blatant example is when a dataset has all its malicious samples originating from the same IP address. If the IP address is considered as a feature, the ML model will only look for the `malicious' IP address, meaning that any attack involving other machines will never be detected.}. If not sanitized, such artifacts may be learned by the ML model to perform its decisions, leading to overfitting and, hence, useless ML-NIDS. 
    \item \textit{Preserving Performance.} When a given context involves modifications of $T$, it is important not to degrade the baseline False Positive Rate (FPR). Modifications of $T$ must always be assessed. 
    \item \textit{Maximizing Performance.} Assuming that simply adding malicious samples to $T$ results in a ML-NIDS capable of detecting such attacks is misleading: it has been shown that ML models for NIDS may yield underwhelming detection performance in multi-classification settings~\cite{Apruzzese:Deep}. It is hence crucial to consider a ML-NIDS architecture that optimizes the usage of such additional samples.
\end{enumerate}
Finally, we highlight three intrinsic \textit{risks} that involve security aspects of cross-evaluations of ML-NIDS.
\begin{itemize}
    \item \textit{Labeling quality.} Cross-evaluations are significant only if the samples in $\mathbb{N}$ and in $\mathbb{M}$ all report the correct label (i.e., benign or malicious). If, e.g., $\mathbb{N}$ contains malicious samples because the authors of the source dataset did not perform proper verifications, then the final results may be unreliable. Unless the cross-evaluation involves unsupervised ML, real deployments should ensure that all samples are associated to the correct ground truth.
    
    \item \textit{Exposure to adversarial ML attacks.} Although mixing data from different networks can result in more resilient ML-NIDS (cf. §\ref{ssec:motivation}), relying on \textit{public} datasets exposes to `poisoning' attacks~\cite{apruzzese2021modeling}. In these circumstances, training a ML-NIDS on such data would have the opposite effect of adversarial training.
    For instance, in~\cite{dunn2020robustness} the FPR increases by 5 times when only 5\% of the data is polluted. More subtle poisoning strategies exploit `backdoors' which make ML-NIDS prone to evasion, as evidenced in~\cite{bachl2019walling}---and also in~\cite{nguyen2020poisoning} for federated learning scenarios. Countermeasures include verifying the checksum of each dataset as provided by the authors; or applying some modifications that can remove or mitigate the effects of such poisoned samples (e.g.,~\cite{Apruzzese:Addressing}).
    
    \item \textit{Incompatible Networks.} Regardless of the resulting performance, mixing samples from different networks may not be possible a-priori. If the goal is using an $N$ from a different network, it is necessary to conduct \textit{preliminary} analyses ensuring that the two networks are indeed similar. On the other hand, when using $M$ from different networks it is necessary to perform \textit{follow-up} analyses that question the validity of the cross-evaluation results. This is because high detection rates at test-time may lead to a `false sense' of security if the malicious activities depend on the underlying network's behaviour. Such analyses may include comparing the feature importance between different models; or complete sanity checks by deploying the ML-NIDS against true attacks--in real time.
    Regardless, using the same source of benign samples both in $T$ and $E$ ensures that the FPR after deployment will not deviate from the one at test-time.
\end{itemize}
We make a crucial remark. Our cross-evaluation idea assumes that $T$ and $E$ always use benign data originating from the same network (i.e., $o$), which is in stark contrast with the practice of `transferring ML models' (common in Computer Vision~\cite{oliver2018realistic}). Indeed, we advise \textit{against} using such practice for ML-NIDS, due to the immense variability of each network~\cite{Sommer:Outside}.

We can conclude that the additional context types enabled by cross-evaluations of ML-NIDS are intriguing, but practical applications are not simple and require the adoption of a rigorous workflow.

%% file: sections/tab_contexts.tex
\begin{table*}[htbp]
\centering
\caption{The 10 types of contexts enabled by the proposed cross-evaluation model. The table represents the relationship among the unique elements of $\mathbb{N}$ and $\mathbb{M}$ included in the training and evaluation sets, $T$ and $E$ (cf. Expr.~\ref{exp:TE}), which denote a context (cf. Expr.~\ref{exp:context}). The origin of benign samples is always the same in $T$ and $E$, hence $\bar{o}$ is shared. Contexts of type \captionC{1}, \captionC{2} are those considered by most prior works; \captionC{4} has been considered in~\cite{pontes2021potts}. A gray background denotes the types of context assessed in our demonstration. \textbf{Example:} consider the setting described in the example of Fig.~\ref{fig:DNM}, where $n$=3 and $\mu$=3. Suppose a context denoted by: $\vv{o}$=(3), $\vv{t}$=(2,3), $\vv{e}$=(3), $\vv{\tau}$=(1,1), $\vv{\varepsilon}$=(1). Such context implies that $T$ contains benign samples from $N_3$, and malicious samples from $M_2^1$ and $M_3^1$; whereas $V$ contains benign samples from $N_3$ but malicious samples from $M_3^1$. The resulting context will be of type \captionC{5}, because $\bar{o}$=(3), $\bar{t}$=(2,3), $\bar{e}$=(3), $\bar{\tau}$=(1), $\bar{\varepsilon}$=(1). Hence, $(\bar{o}=\bar{e})\neq\bar{t}$ and $\bar{\tau}=\bar{\varepsilon}$.} 
    \resizebox{2\columnwidth}{!}{
        \begin{tabular}{c?c|c?c}
            \toprule
             \textbf{\C{}-type} & \multicolumn{1}{c|}{\textbf{Conditions}} & \textbf{Effects on $T$ and $E$} & \textbf{Use-case} \\

             \toprule
            \rowcolor{gray!15}\C{1} &
            
            \begin{tabular}{c}
                \condeq{o}{t} \\
                \condeq{o}{e} \\
                \condeq{t}{e} \\
                \condeq{\tau}{\varepsilon} \\
            \end{tabular} &
            \begin{tabular}{c}
                $T$ uses benign and malicious samples from the same network. \\
                $E$ uses benign and malicious samples from the same network. \\
                $T$ and $E$ use malicious samples from the same network.\\
                $T$ and $E$ use the same attack classes.\\
            \end{tabular} & 
            \begin{tabular}{c}
                This is the standard `baseline' case \\
                commonly used in research. \\ 
            \end{tabular} \\
            \midrule
            
            \C{2} &
            \begin{tabular}{c}
                \condeq{o}{t} \\
                \condeq{o}{e} \\
                \condeq{t}{e} \\
                \condneq{\tau}{\varepsilon} \\
            \end{tabular} & 
            \begin{tabular}{c}
                $T$ uses benign and malicious samples from the same network.\\
                $E$ uses benign and malicious samples from the same network.\\
                $T$ and $E$ use malicious samples from the same network.\\
                $T$ and $E$ use different attack classes.\\
            \end{tabular}  & 
            \begin{tabular}{c}
                Same as the baseline \C{1} \\
                but the ML-NIDS is tested on different attacks. \\ 
            \end{tabular} \\
            
            \midrule
            
            \C{3} &
            \begin{tabular}{c}
                \condeq{o}{t} \\
                \condneq{o}{e} \\
                \condneq{t}{e} \\
                \condeq{\tau}{\varepsilon} \\
            \end{tabular} & 
            \begin{tabular}{c}
                $T$ uses benign and malicious samples from the same network.\\
                $E$ uses benign and malicious samples from different networks.\\
                $T$ and $E$ use malicious samples from different networks.\\
                $T$ and $E$ use the same attack classes.\\
            \end{tabular}  & 
            \begin{tabular}{c}
                Using the ML-NIDS on the same attacks,\\
                but targeting hosts from a different network. \\ 
            \end{tabular} \\
            \midrule
            
            \rowcolor{gray!15}\C{4} &
            \begin{tabular}{c}
                \condeq{o}{t} \\
                \condneq{o}{e} \\
                \condneq{t}{e} \\
                \condneq{\tau}{\varepsilon} \\
            \end{tabular} & 
            \begin{tabular}{c}
                $T$ uses benign and malicious samples from the same network.\\
                $E$ uses benign and malicious samples from different networks.\\
                $T$ and $E$ use malicious samples from different networks.\\
                $T$ and $E$ use different attack classes.\\
                
            \end{tabular}  & 
            \begin{tabular}{c}
                Testing the generalization capabilities of the NIDS \\ 
                on completely unknown attacks  \\
                (different hosts, different attacks).
            \end{tabular} \\
            \midrule
            
            \C{5} &
            \begin{tabular}{c}
                \condneq{o}{t} \\
                \condeq{o}{e} \\
                \condneq{t}{e} \\
                \condeq{\tau}{\varepsilon} \\
            \end{tabular} & 
            \begin{tabular}{c}
                $T$ uses benign and malicious samples from different networks.\\
                $E$ uses benign and malicious samples from the same network.\\
                $T$ and $E$ use malicious samples from different networks.\\
                $T$ and $E$ use the same attack classes.\\
            \end{tabular}  & 
            \begin{tabular}{c}
                When there are too few samples of an attack class\\to both train and test a ML-NIDS,\\it is possible to borrow some malicious\\samples from another network and use them in $T$.
            \end{tabular} \\
            \midrule
            
            \C{6} &
            \begin{tabular}{c}
                \condneq{o}{t} \\
                \condeq{o}{e} \\
                \condneq{t}{e} \\
                \condneq{\tau}{\varepsilon} \\
            \end{tabular} &
            \begin{tabular}{c}
                $T$ uses benign and malicious samples from different networks.\\
                $E$ uses benign and malicious samples from the same network.\\
                $T$ and $E$ use malicious samples from different networks.\\
                $T$ and $E$ use different attack classes.\\
            \end{tabular}  & 
            \begin{tabular}{c}
                Assessing whether training on new attacks \\
                (different hosts and attack class) \\
                affects the detection on `known' attacks.
            \end{tabular} \\
            \midrule
            
            \rowcolor{gray!15}\C{7} &
            \begin{tabular}{c}
                \condneq{o}{t} \\
                \condneq{o}{e} \\
                \condeq{t}{e} \\
                \condeq{\tau}{\varepsilon} \\
            \end{tabular} & 
            \begin{tabular}{c}
                $T$ uses benign and malicious samples from different networks.\\
                $E$ uses benign and malicious samples from different networks.\\
                $T$ and $E$ use malicious samples from the same network.\\
                $T$ and $E$ use the same attack classes.\\
            \end{tabular}  & 
            \begin{tabular}{c}
                Extending the detection surface of the ML-NIDS\\by training on attacks originating from different networks\\and testing such ML-NIDS against these attacks.\\
            \end{tabular} \\
            \midrule
            
            \C{8} &
            \begin{tabular}{c}
                \condneq{o}{t} \\
                \condneq{o}{e} \\
                \condeq{t}{e} \\
                \condneq{\tau}{\varepsilon} \\
            \end{tabular} & 
            \begin{tabular}{c}
                $T$ uses benign and malicious samples from different networks.\\
                $E$ uses benign and malicious samples from different networks.\\
                $T$ and $E$ use malicious samples from the same network.\\
                $T$ and $E$ use the different attack classes.\\
            \end{tabular}  & 
            \begin{tabular}{c}
                Testing an `extended' ML-NIDS devised by exploiting \C{7} \\
                against other attacks for which not enough samples \\
                are available to train a dedicated detector.\\
            \end{tabular} \\
            \midrule
            
            \C{9} &
            \begin{tabular}{c}
                \condneq{o}{t} \\
                \condneq{o}{e} \\
                \condneq{t}{e} \\
                \condeq{\tau}{\varepsilon} \\
            \end{tabular} & 
            \begin{tabular}{c}
                $T$ uses benign and malicious samples from different networks.\\
                $E$ uses benign and malicious samples from different networks.\\
                $T$ and $E$ use malicious samples from different networks.\\
                $T$ and $E$ use the same attack classes.\\
            \end{tabular}  & 
            \begin{tabular}{c}
                Using the benign samples of a different network \\ 
                and combining them with `owned' malicious samples \\
                whether the `owned' malicious samples\\
                
            \end{tabular} \\
            \midrule
            
            \C{10} &
            \begin{tabular}{c}
                \condneq{o}{t} \\
                \condneq{o}{e} \\
                \condneq{t}{e} \\
                \condneq{\tau}{\varepsilon} \\
            \end{tabular} & 
            \begin{tabular}{c}
                $T$ uses benign and malicious samples from different networks.\\
                $E$ uses benign and malicious samples from different networks.\\
                $T$ and $E$ use malicious samples from different networks.\\
                $T$ and $E$ use different attack classes.\\
            \end{tabular}  & 
            \begin{tabular}{c}
                Training a ML-NIDS using benign and malicious sample\\from two distinct subnets, and testing the generalization\\ capabilities of such ML-NIDS against\\other attacks targeting a different subnet.
                
            \end{tabular} \\
            
            \bottomrule
        
        \end{tabular}
    }
\label{tab:contexts}
\end{table*}

%% file: sections/4-framework.tex
\section{Proposed Framework: XeNIDS}
\label{sec:framework}
We showed that cross-evaluations of ML-NIDS are enticing but challenging, and we are not aware of efforts that tackled this problem in an exhaustive way. 
As a first step, we propose XeNIDS, a framework for the Cross-evaluation of Network Intrusion Detection Systems based on machine learning, with a focus on NetFlow data. 
Our proposed framework is rooted in the same design principles described in §\ref{ssec:principles}, and has a threefold goal:
\begin{itemize}
    \item allowing the \textit{simulation} of all contexts in Table~\ref{tab:contexts};
    \item facilitating assessments of \textit{multiple} contexts;
    \item addressing the challenges discussed in §\ref{ssec:challenges}.
\end{itemize}
Of course, we do not claim that XeNIDS is the only way to do all of the above. Our intention is to further promote the diffusion of cross-evaluations in \textit{research}, as well as to increase their realistic value for proactive assessments.

We provide an overview of XeNIDS in §\ref{ssec:overview}, which consists in four stages: \textit{standardize} (§\ref{ssec:standardize}), \textit{isolate} (§\ref{ssec:isolate}), \textit{contextualize} (§\ref{ssec:contextualize}), \textit{cross-evaluate} (§\ref{ssec:cross-evaluate}). 

\subsection{Overview}
\label{ssec:overview}
The focus of XeNIDS is on NetFlows (§\ref{ssec:related}), enabling \textit{inter-compatibility} with PCAP data. NetFlows are metadata generated from packet captures, and summarize the communications between two endpoints. A NetFlow is defined as:
\begin{equation}
\label{exp:flow}
\resizebox{0.8\columnwidth}{!}{
$
    \text{NetFlow}\!=\!{\text{\small(\textit{srcIP}, \textit{dstIP}, \textit{srcPort}, \textit{dstPort}, \textit{t}, \textit{proto}, \textit{d}, ...,)}}
    $,
    }
\end{equation}
where \textit{srcIP} (\textit{srcPort}) and \textit{dstIP} (\textit{dstPort}) are the source and destination IP addresses (network ports) of the two involved hosts, \textit{t} is the timestamp of the first connection, \textit{d} is the duration of the communication session, \textit{proto} is the network protocol of the communication. Depending on the NetFlow software and its configuration, additional metrics can be computed: the most typical fields include the  number of \textit{packets} and \textit{bytes} exchanged during the communication~\cite{vormayr2020my}. 

\begin{figure}[!htbp]
    \centering
    \includegraphics[width=\columnwidth]{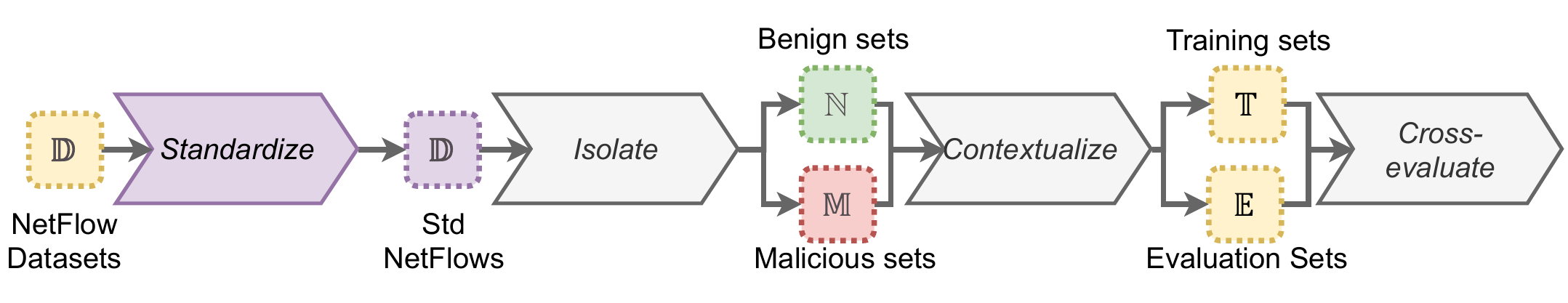}
    \caption{Overview of XeNIDS. The input $\mathbb{D}$ is a set of labelled NetFlow datasets of $n$ distinct networks. The output results should be further analyzed.}
    \label{fig:overview}
\end{figure}

We present a schematic representation of XeNIDS in Fig.~\ref{fig:overview}.

XeNIDS requires a set of $n$ datasets containing NetFlows, representing $\mathbb{D}$ and totalling $\mu$ distinct attack classes. 

These datasets must be provided with the \textit{ground truth}. XeNIDS assumes that all data in $\mathbb{D}$ is verified, trusted and appropriate for the considered deployment scenario (§\ref{ssec:challenges}).
 
The framework includes four stages (cf. Fig.~\ref{fig:overview}):
\begin{enumerate}
    \item \textit{Standardize}: the input datasets in $\mathbb{D}$ are first cleaned and sanitized, and then brought into a common `language'.
    \item \textit{Isolate}: every standardized dataset is partitioned in its benign and malicious sets ($\mathbb{N}$ and $\mathbb{M}$).
    \item \textit{Contextualize}: $\mathbb{M}$ and $\mathbb{N}$ are used to compose a context (by generating the corresponding $\mathbb{T}$ and $\mathbb{V}$).
    \item \textit{Cross-evaluate}: $\mathbb{T}$ and $\mathbb{V}$ are used to develop and cross-evaluate a ML-NIDS.
\end{enumerate}

The results provided as \textit{output} by XeNIDS should be \textit{further analyzed} for practical deployments (cf. §\ref{ssec:challenges}).

\subsection{Standardize}
\label{ssec:standardize}
\begin{figure}[!htbp]
    \centering
    \includegraphics[width=\columnwidth]{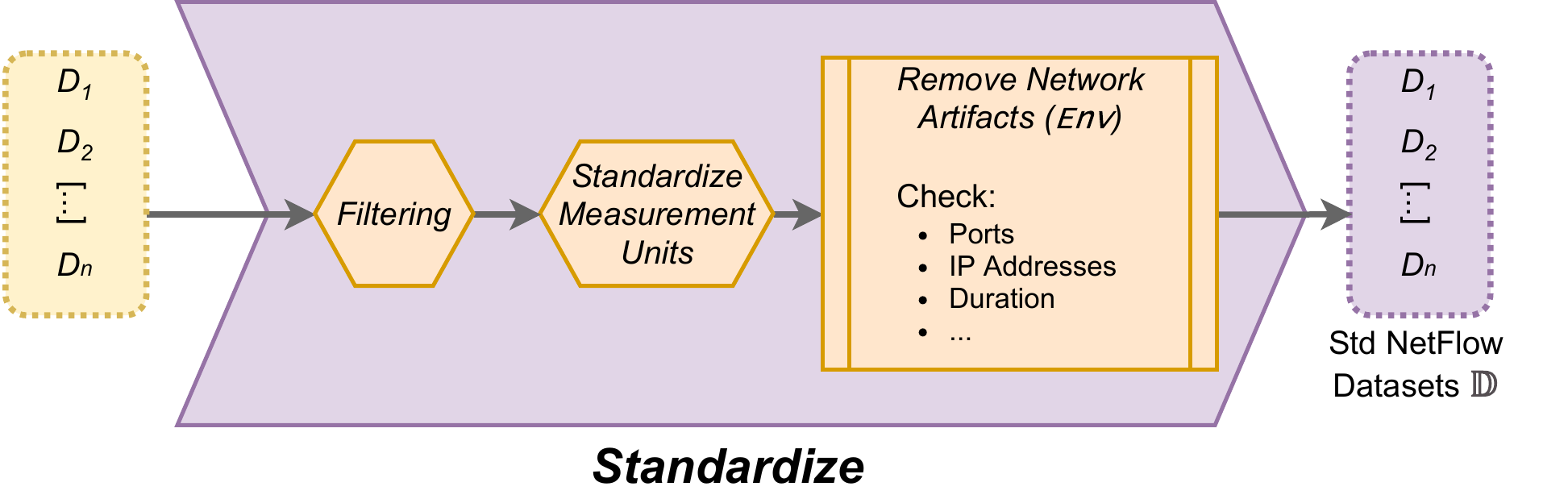}
    \caption{First stage: Standardize. The initial NetFlows in $\mathbb{D}$ are all standardized to derive a common feature set, and cleaned of any possible artifact that may lead to overfitting and impractical ML-NIDS.}
    \label{fig:standardization}
\end{figure}

In the first stage, schematically depicted in Fig.~\ref{fig:standardization}, XeNIDS brings all the datasets $D_i \in \mathbb{D}$ into a common NetFlow format, accounting for potential obfuscations as a result of \textit{anonymization} techniques. Essential operations involve data sanitization (e.g., handling missing values) and filtering: for example, if the goal is the detection of attacks involving TCP traffic, then all non-TCP traffic can be safely removed. Then, the focus is on establishing a common feature set\footnote{Taken from the \textit{intersection} of the features across all $\mathbb{D}$.} while simultaneously \textit{removing network artifacts} that may lead to overfitting. Such procedures are tough, especially when considering NetFlow records, but necessary. To explain the reasons of such difficulties and our proposed workarounds, we provide an \textit{original interpretation} of NetFlows with respect to machine learning.

In simple terms, a NetFlow is the result of two contributors: the \textit{communications} ($Comm$) performed by the involved hosts, and the effects of the \textit{environment} ($Env$) where the NetFlow is generated. This latter factor ($Env$) is, in turn, influenced by two elements: the network \textit{identity} ($NetId$), denoting the intrinsic characteristics of the network where $Comm$ (such as allocated bandwidth, protocols used, common open ports, periodic services) are captured; and the \textit{configuration} of the appliance ($Conf$) used to generate the NetFlows. 
Hence, the information captured by a NetFlow is a function\footnote{The definition of $\mathcal{F}$ is software dependent~\cite{vormayr2020my}, and outside our scope.} $\mathcal{F}$ of three components: $Comm$, $NetId$, $Conf$. Formally:

 \begin{align}
    \resizebox{0.75\columnwidth}{!}{
  $\begin{aligned}
      \mathcal{F}(Comm, Env) \Rightarrow \mathcal{F}(Comm, NetId, Conf) = \text{NetFlow}
  \end{aligned}$
  }
  \label{exp:env}
\end{align} 

\par
The ultimate goal of the standardize stage is to mitigate the effects of $Env$ (represented by $NetId$ and $Conf$) across all the input datasets in $\mathbb{D}$. Indeed, if one dataset $D_i$ has an $Env$ that is significantly stronger than $D_j$, then a ML model trained on data from $D_i$ and $D_j$ may only learn on the basis of such `signature' $Env$. These circumstances lead to overfitting on $Env$, resulting in impractical detectors that neglect to search for malicious behaviours.

Let us explain Exp.~\ref{exp:env} with two practical use-cases on the contribution of $Env$. 
\begin{itemize}
    \item Different $NetId$. Consider two different networks where a host downloads the same file from the same remote server via SSH: if these two networks use different listening ports for the SSH server (e.g., 22 and 4022), then the NetFlows of the first network will differ from those of the second network (they will have different ports). 
    
    \item Different $Conf$. Using different NetFlow software and/or settings yields different NetFlows even when the original PCAP traces are identical. For instance, measurement units can differ, resulting in datasets that are not compatible: a dataset $D_i$ with $d$ expressed in milliseconds cannot be used alongside a dataset $D_j$ that uses seconds.
\end{itemize}
We report in Appendix~\ref{app:influence} an exhaustive explanation of the effects brought by $Env$ on NetFlows.

By referring to the official NetFlow v9 documentation\footnote{\url{www.cisco.com/c/en/us/products/ios-nx-os-software/ios-netflow/}}, we observe that there are several fields that can contribute to $Env$ (influenced both by $NetId$ and $Conf$), which require particular care at this stage. We provide in Appendix~\ref{app:recommendations} some recommendations for reducing the generation of the above-mentioned artifacts, with a focus on three fields: the IP address, the network ports, and the flow duration. 

Nonetheless, depending on the considered use-cases, many low-level implementations are viable to minimize the impact of $Env$ and derive a common feature set.
After this stage, the initial set of datasets $\mathbb{D}$ is standardized and ready for the `core' functionalities of XeNIDS.

\subsection{Isolate}
\label{ssec:isolate}
\begin{figure}[!htbp]
    \centering
    \includegraphics[width=\columnwidth]{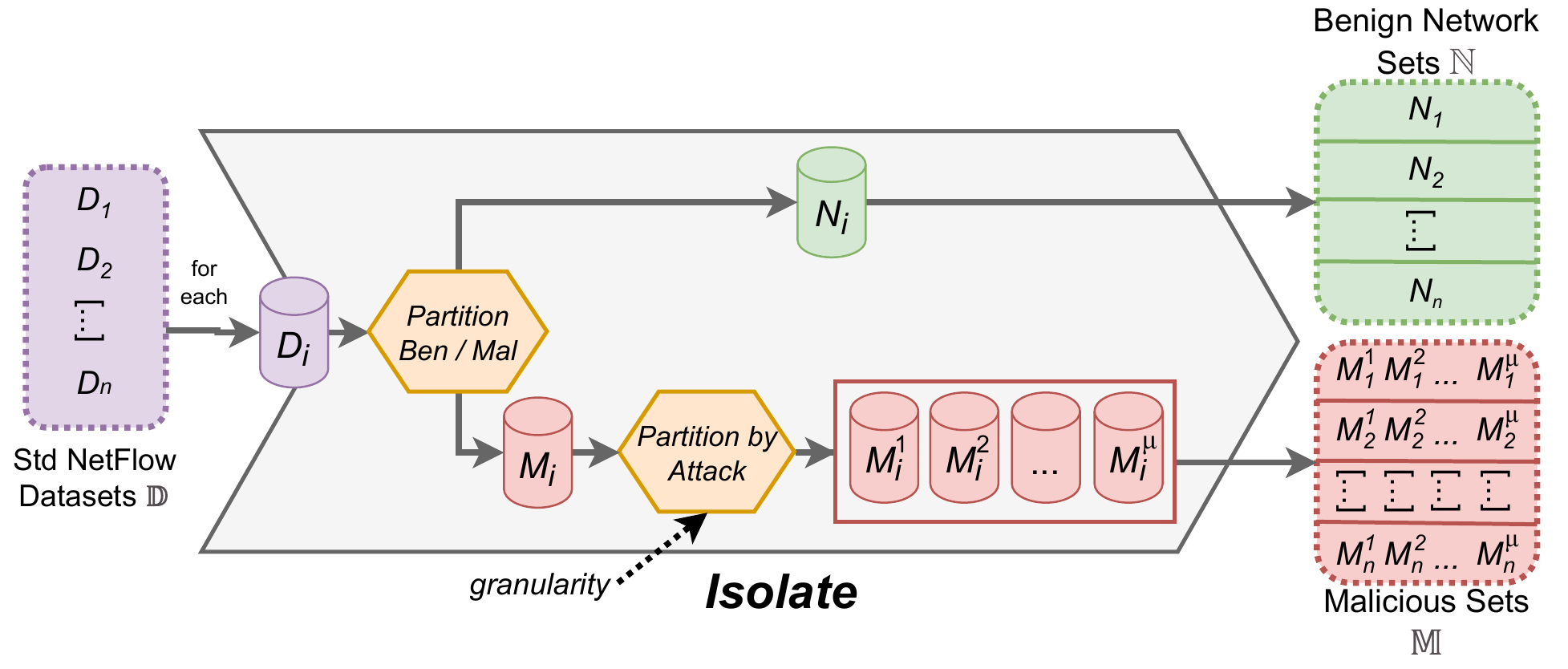}
    \caption{Second stage: isolate. The standardized $\mathbb{D}$ is used to extract the $n$ individual benign $N$ and malicious $M$ sets. The latter are then split in $\mu$ sets on the basis of their attack class. Depending on the user-provided granularity, it is possible to aggregate sets of different attack classes into a bigger class, therefore changing the initial $\mu$. The outputs are the full sets $\mathbb{N}$ and $\mathbb{M}$.}
    \label{fig:isolation}
\end{figure}

In this stage, XeNIDS isolates the benign from malicious samples of each (standardized) dataset in $\mathbb{D}$ to derive $\mathbb{N}$ and $\mathbb{M}$ (cf. Fig.~\ref{fig:DNM} in §\ref{ssec:principles}). We provide a schematic in Fig.~\ref{fig:isolation}.

Specifically, XeNIDS first partitions the benign from malicious samples in each $D_i$, resulting in two distinct sets, $N_i$ and $M_i$. Then, XeNIDS further partitions the specific attack samples in $M_i$ according to the \textit{individual} attack that they represent\footnote{The attack is denoted by the ground-truth labels provided in the input $\mathbb{D}$.} (assuming that $\mu > 1$). 

Such design choice enables the development of \textit{collaborative ensembles} (e.g.,~\cite{pontes2021potts, acharya2021efficacy}) of ML classifiers, each devoted to a specific threat, therefore addressing the third challenge (cf. §\ref{ssec:challenges})---while also allowing to use XeNIDS for multi-classification ML problems.

We note that the separation can also account for a specific level of \textit{granularity}. In this case, the original $\mu$ will be changed by `aggregating' attacks of different classes, potentially even treating all malicious samples as belonging to a single malicious class. Depending on the use-case, such granularity can vary: it could either be performed at a high-level (e.g., \textit{Botnet} or \textit{DoS} attacks) or go at a deeper level (e.g., a specific \textit{Botnet} variant). The final choice depends on the actual use-case (e.g., when there are not enough samples available, they can be aggregated into a macro-class, or simply discarded). 

This stage produces two outputs: $\mathbb{N}$, containing all the benign network samples (partitioned in $n$ subsets according to their source dataset); 
and $\mathbb{M}$, containing all malicious samples isolated in $n \times \mu$ subsets of samples according to their specific attack and source dataset. We recall that some elements of $\mathbb{M}$ can be empty, i.e., if a $D_i$ does not contain malicious samples of the same classes as $D_j$. 

\subsection{Contextualize}
\label{ssec:contextualize}
\begin{figure}[!htbp]
    \centering
    \includegraphics[width=\columnwidth]{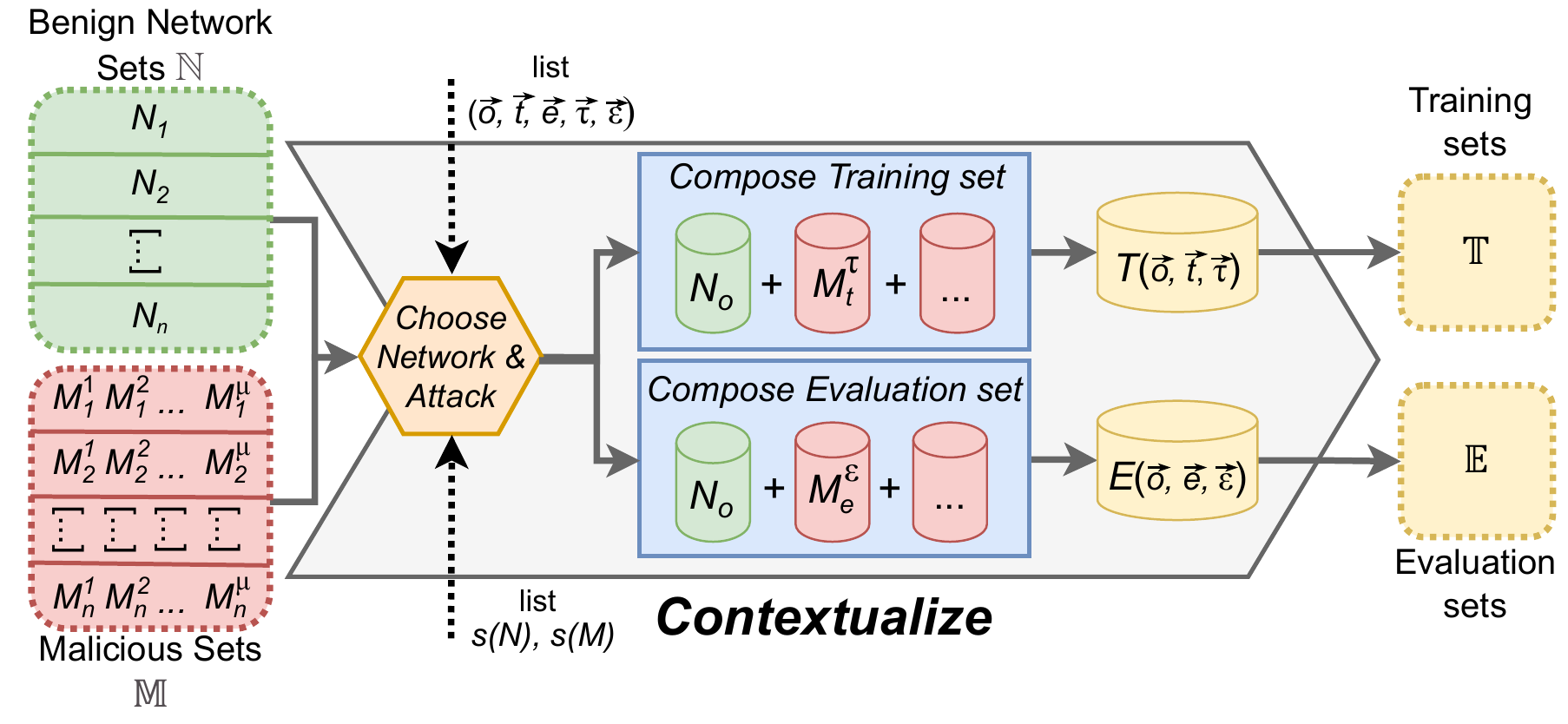}
    \caption{Third stage: contextualize. XeNIDS uses $\mathbb{N}$ and $\mathbb{M}$ to create multiple $T$ and $E$, according to the context-related parameters ($\vv{o}$, $\vv{t}$, $\vv{e}$, $\vv{\tau}$, $\vv{\varepsilon}$)) and the splits $s(N)$ and $s(M)$. All the composed $T$ and $E$ are inserted in $\mathbb{T}$ and $\mathbb{E}$.
    \textbf{Example.} Assume that the following user-provided parameters: $\vv{o}$=(1), $\vv{t}$=(2), $\vv{e}$=(3), $\vv{\tau}$=(1), $\vv{\varepsilon}$=(1); $s(N)$=(80:20), $s(M)$=(70:30). XeNIDS chooses $N_1$, and puts 80\% of $N_1$ in $T$ and the remaining 20\% in $V$. XeNIDS then selects $M_2^1$ and puts 70\% of its samples in $T$; then XeNIDS selects $M_3^1$ and puts 30\% of its samples in $V$. Such $T$ (and $V$) is then inserted in $\mathbb{T}$ (and $\mathbb{V}$). The operation is repeated if the user provides additional lists as input.}
    \label{fig:contextualize}
\end{figure}

In the third stage, XeNIDS creates \textit{all} the sets corresponding to the contexts to simulate during the cross-evaluation. This is done by using $\mathbb{N}$ and $\mathbb{M}$ (provided by the previous stage), alongside some external input, to compose training $T$ and evaluation $E$ sets; all such $T$ and $E$ will be put in two dedicated collections, $\mathbb{T}$ and $\mathbb{E}$. We provide a schematic of this stage in Fig.~\ref{fig:contextualize}.

Two user-provided input \textit{lists} regulate this stage: a 5-dimensional tuple of context-related parameters ($\vv{o}$, $\vv{t}$, $\vv{e}$, $\vv{\tau}$, $\vv{\varepsilon}$); and a pair of \textit{splits}, $s(N)$ and $s(M)$, used to partition (e.g., 80:20) any $N$ and $M$ that will be included in a given $T$ and $E$. 
The idea is to facilitate cross-evaluations that consider multiple contexts, by composing all the necessary $T$ and $E$ \textit{before} using them for any assessment. Hence, XeNIDS iterates over all the elements in the two input lists: at each iteration, XeNIDS composes a training and evaluation set according to the user-specified parameters.

Specifically, for each tuple ($\vv{o}$, $\vv{t}$, $\vv{e}$, $\vv{\tau}$, $\vv{\varepsilon}$), and for each pair of splits, $s(N)$ and $s(M)$, XeNIDS proceeds as follows.
\begin{itemize}
    \item XeNIDS uses $o$ to select a specific set of benign samples $N_o$ from $\mathbb{N}$. XeNIDS splits $N_o$ according to $s(N)$, and puts the corresponding partitions in $T$ and $E$.
    \item For each $(t,\tau)\!\!\in\!\!(\vv{t},\vv{\tau})$, XeNIDS extracts from $\mathbb{M}$ the element $M_{t}^{\tau}$, which is split with $s(M)$ and put in $T$.
    \item For each $(e,\varepsilon)\!\in\!(\vv{e},\vv{\varepsilon})$, XeNIDS extracts from $\mathbb{M}$ the element $M_{e}^{\varepsilon}$, which is split with $s(M)$ and put in $E$.
    \item Alltogether, these operations result in two sets, $T$($\vv{o}$,$\vv{t}$,$\vv{\tau}$) and $E$($\vv{o}$,$\vv{e}$,$\vv{\varepsilon}$), which are put in $\mathbb{T}$ and $\mathbb{E}$. 
\end{itemize}
An example of such workflow is in the caption of Fig.~\ref{fig:contextualize}.

In cases where $t$=$e$ and $\tau$=$\varepsilon$, XeNIDS performs the partitioning simultaneously, to avoid overlaps that can result in the same malicious samples being included in both $T$ and $E$. 

The selection of $s(N)$ and $s(M)$, which can differ for $T$ and $E$, must be done to achieve a twofold goal: (i) realize an $E$ that is comprehensive enough to cover the real data distribution, and hence produce insightful results; and (ii) realize a $T$ that allows to develop proficient ML-NIDS.
For instance, if $E$ does not contain many benign samples, then the resulting FPR may not correspond to the real FPR after the ML-NIDS is deployed, At the same time, if $T$ contains only a small number of samples for a given attack, the resulting ML-NIDS will not be able to capture all the possible variations of such attack\footnote{We refer the reader to~\cite{zhang2021network} for a study on how the size of the training set can impact the performance of ML-NIDS.}.

After this stage, we obtain two collections of training and evaluation sets, $\mathbb{T}$ and $\mathbb{E}$. 

\subsection{Cross-evaluate}
\label{ssec:cross-evaluate}
In the last stage, XeNIDS performs the cross-evaluation by using the sets in $\mathbb{T}$ and $\mathbb{E}$ to reproduce any user-specified context. Hence, the input parameters are a list of contexts \C~(cf. Exp.~\ref{exp:context}); as well as a learning ML algorithm to develop the detectors of the ML-NIDS. 

\begin{figure}[!htbp]
    \centering
    \includegraphics[width=\columnwidth]{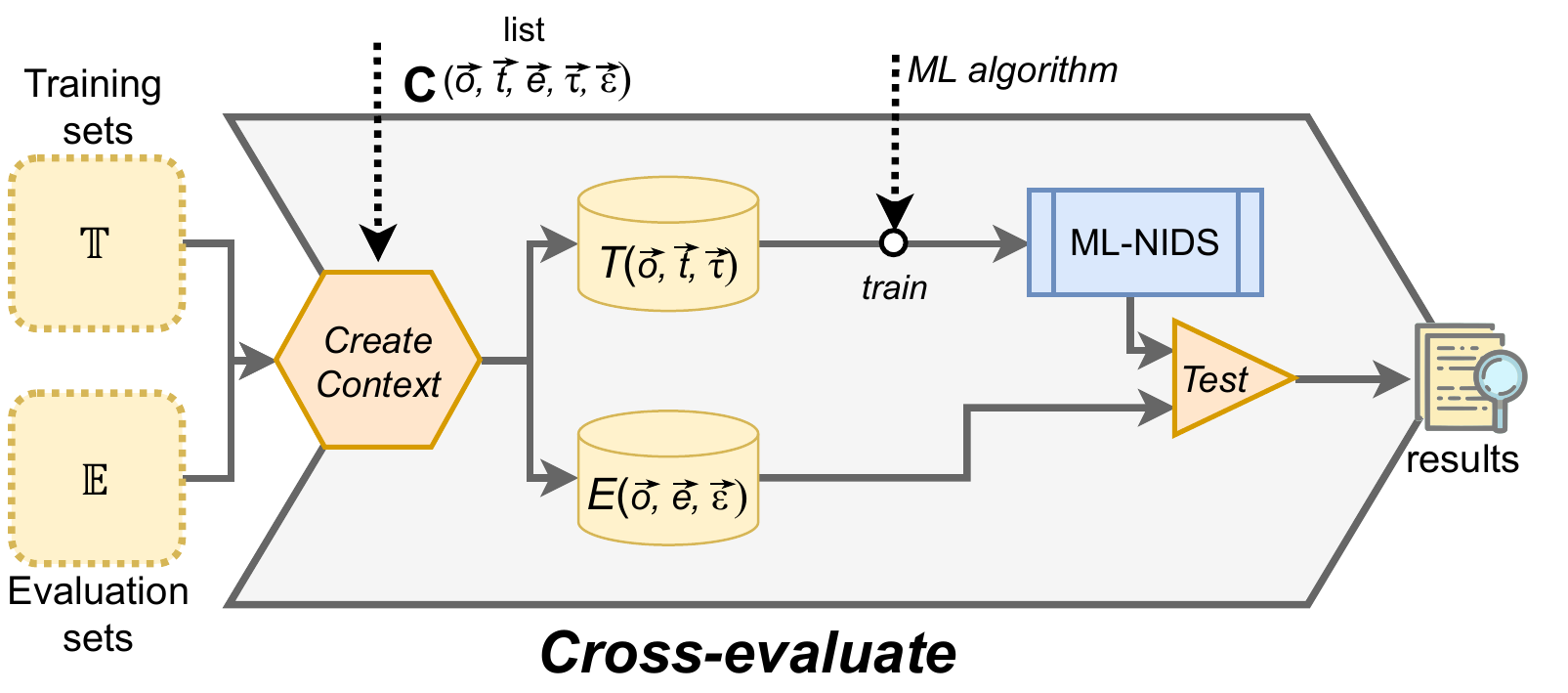}
    \caption{Final stage: cross-evaluate. XeNIDS reproduces the user-specified context(s) and performs the cross-evaluation.  \textbf{Example:} assume the following input context: \captionC{}((1), (1,2), (1,2), (1,1), (2,2)) and a ML-NIDS that leverages ensembles of binary detectors. XeNIDS first extracts $T$((1),(1,2),(1,1)) from $\mathbb{T}$, which is split in two smaller sets, $T$(1,1,1) and $T$(1,2,1). Such sets are used to train two ML-models that will compose the ML-NIDS. The ML-NIDS can then be tested either against $E$((1),(1,2),(2,2)), or against its subsets $E$(1,1,2) and $E$(1,2,2), all of which obtained from $\mathbb{E}$. Any previously trained model (e.g., the one using $T$(1,1,1)) can be reused to assess different contexts.} 
    \label{fig:cross-evaluate}
\end{figure}

Specifically, for each context \C(\smallarray{o}, \smallarray{t}, \smallarray{e}, \smallarray{\tau} provided as input, XeNIDS draws the corresponding $T$(\smallarray{o}, \smallarray{t}, \smallarray{\tau}) and $E$(\smallarray{o}, \smallarray{e}, \smallarray{\varepsilon}) from $\mathbb{T}$ and $\mathbb{E}$. Then, depending on the architecture of the ML-NIDS, XeNIDS operates as follows.
\begin{itemize}
    \item If the ML-NIDS leverages a \textit{single classifier}, XeNIDS uses $T$(\smallarray{o}, \smallarray{t}, \smallarray{\tau}) to train a single (multi-class) ML-model with a given ML algorithm; such ML-model is then tested against $E$(\smallarray{o}, \smallarray{e}, \smallarray{\varepsilon}).
    \item If the ML-NIDS leverages \textit{ensembles of classifiers}, XeNIDS splits $T$(\smallarray{o}, \smallarray{t}, \smallarray{\tau}) into smaller sets, (e.g., by composing $T$($o$,$t$,$\tau$) focusing on the specific attack $\tau$ contained in $t$); each of these sets is used to train a dedicated ML-model of the ensemble. Such procedure can be repeated for $E$(\smallarray{o}, \smallarray{e}, \smallarray{\varepsilon}), i.e., the ML-NIDS can be tested against the entire $E$, or against subsets.
\end{itemize}
The design of XeNIDS enables the assessment of multiple contexts without the need of training additional ML models. If two contexts require the same $T$, it is only necessary to draw a different $E$ from $\mathbb{E}$, and use such $E$ to assess the previously trained ML-NIDS.

We illustrate this stage in Fig.~\ref{fig:contextualize}, where we also provide a complete example of an ensemble use-case. We anticipate that, in our demonstration, we will always use ensembles of specialized classifiers.

The results produced as output of this stage should be subject to subsequent analyses and considerations.

%% file: sections/5-application.tex
\section{Application}
\label{sec:application}
As a final contribution of this paper, we showcase\footnote{Our implementation of XeNIDS: \url{https://github.com/pajola/XeNIDS}} a practical application of XeNIDS. We do so via a large set of experiments where we cross-evaluate ML-NIDS by using a total of 6 well-known NID datasets. 
We describe our testbed (§\ref{ssec:datasets}) and explain the preprocessing operations (§\ref{ssec:preprocessing}). Then, we present the common assessment procedure (§\ref{ssec:assessment}). 

\subsection{Testbed}
\label{ssec:datasets}
The aim of our demonstration is reproducing and assessing the use-cases described in §\ref{sssec:usecase}. To this purpose, we assess three different context types (cf. Table~\ref{tab:contexts}), namely \C{1}, \C{4} and \C{7}. However, we differentiate our experiments depending on the \textit{format} of the NetFlow data used as input to XeNIDS: specifically, such data can be either in a \textit{uniform} or \textit{heterogeneous} format. Let us explain our rationale and the differences between these two distinct scenarios.

We recall that XeNIDS operates on \textit{existing} data in the form of NetFlows. Such NetFlows can be provided either (a) as PCAP traces, and then exported to NetFlows using dedicated software; or (b) directly as NetFlows, processed according to the creators' specifications.
These two scenarios must be treated separately, due to the different effects that they can have on the \textit{results}. In the first scenario, the raw PCAP traces (collected in diverse network environments) can be used to generate \textit{uniform} NetFlows by using the same appliance for all PCAP traces; because the NetFlows share the same format,  the results are more reliable due to a lower chance of network artifacts (the contribution of $Conf$ is the same---cf. Exp.~\ref{exp:env}).
However, such scenario requires \textit{all} source data to be fully provided as PCAP, which is a requirement that is hard to meet\footnote{E.g., PCAP data can be truncated~\cite{Garcia:CTU} or not fully labelled~\cite{CICDDOS2019:Dataset}.}. Therefore, it is insightful to consider also the scenario where the source datasets are provided directly as \textit{heterogeneous} NetFlows (due to being generated with different software). Such scenario requires a more careful application of XeNIDS's \textit{standardize} stage (§\ref{ssec:standardize}), but also a more detailed analysis of the results because the effects of different initial $Conf$ can only be seen \textit{after} the ML-NIDS is evaluated.

We hence apply XeNIDS differently for both scenarios, each considering 4 well-known datasets ($n$=4 for both scenarios).
\begin{itemize}
    \item \textit{Heterogeneous} scenario: here, we use the \dataset{CTU13}~\cite{Garcia:CTU}, \dataset{NB15}~\cite{UNSWNB15:Dataset}, \dataset{IDS18}~\cite{CICIDS2018:Dataset}, \dataset{DDOS19}~\cite{CICDDOS2019:Dataset}. These are all provided as NetFlows, but using different appliances.
    
    \item \textit{Uniform} scenario: here, we use the \dataset{UF-BotIoT}, \dataset{UF-NB15}, \dataset{UF-IDS18}, \dataset{UF-ToNIoT}. These datasets are created in~\cite{sarhan2020netflow} by the PCAP version of existing datasets and generating the corresponding (labelled) NetFlows using a unified appliance. 
\end{itemize}
For comparison purposes, two datasets are shared\footnote{\captiondataset{UF-NB15} and \captiondataset{UF-IDS18} are generated from \captiondataset{NB15} and \captiondataset{IDS18}.}, whereas two are unique for each scenario.

We provide in Table~\ref{tab:dataset} an overview of these datasets. For each dataset, we report the amount of NetFlows, the overall number of malicious classes, the size of the provided feature-set, and the performance (as F1-score) achieved by the state-of-the-art. From this table, we can already observe the effects of $Conf$ on the corresponding NetFlows: two datasets (i.e., the \dataset{NB15} and the \dataset{IDS18}) are used by both scenario, but the amount of samples and features differ. For example, the \dataset{NB15} has 2.5M samples in the \textit{heterogeneous} scenario, and 1.6M in the \textit{uniform} scenario. Moreover, let us focus on the performance achieved by past works. We can see that the state-of-the-art reaches very high F1-scores, which can raise the question of whether there is any point in improving such values. Nevertheless, we stress that cross-evaluations have a different objective (cf. §\ref{ssec:motivation}): assessing the effectiveness of ML-NIDS against different attacks (\textit{not included} in the respective datasets), and -- if necessary -- improving such ML-NIDS against these attacks. As our results will show, most of these ML-NIDS will perform poorly against different attacks, but can be strengthened; such achievements, however, could only be obtained by cross-evaluations.

\begin{table}[!ht]
    \centering
    \caption{Statistics of the analyzed NID datasets.}
    \resizebox{0.98\columnwidth}{!}{
        \begin{tabular}{c|ccccccc}\toprule
            \textbf{Scenario} & \textbf{Dataset} & \textbf{\#Samples} & \textbf{\#Attacks} & \textbf{\#Features}& \textbf{F1-score}\\ \midrule
            
            \multirow{4}{*}{{
            {\parbox[c]{1.2cm}{{\hspace{3mm}\textit{Heter.}}}}
            }}

            & \dataset{CTU13} & 20.7M  & 5  & 14 & $99.1\%$~\cite{apruzzese2020deep}\\
            & \dataset{NB15} & 2.5M  & 9 & 48 & $98.7\%$~\cite{injadat2020multi}\\
            & \dataset{IDS18}& 3.1M & 14 & 80 & $96.2\%$~\cite{vinayakumar2019deep}\\
            & \dataset{DDOS19} & 70M & 18  & 80 & $99.0\%$~\cite{pontes2021potts}\\ 
            \midrule
            
            \multirow{4}{*}{\textit{Uniform}}

            & \dataset{UF-BotIoT} & 600K  & 4  & 12 & $97.0\%$~\cite{sarhan2020netflow}\\
            & \dataset{UF-NB15} & 1.6M & 9 & 12 & $85.0\%$~\cite{sarhan2020netflow}\\
            & \dataset{UF-IDS18} & 8.3M & 14 & 12 & $83.0\%$~\cite{sarhan2020netflow}\\
            & \dataset{UF-ToNIoT}& 1.4M & 9 & 12 & $100.0\%$~\cite{sarhan2020netflow}\\

            \bottomrule
            
        \end{tabular}
    }
    
    \label{tab:dataset}
\end{table}

Overall, these datasets contain traffic captured in large networks and the included malicious samples belong to a broad range of attacks\footnote{For a precise description of each attack, we refer the reader to the source material provided by the creators of each dataset.}
Table~\ref{tab:dataset2} shows the attack distribution of the input $\mathbb{D}$ for both scenarios. For simplicity, we organize Table~\ref{tab:dataset2} on the basis of three `families' of attacks:
\begin{itemize}
    \item \textit{DoS}, for Denial of Service attacks (e.g., \textit{DoS-Hulk});
    \item \textit{Botnet}, for Botnet attacks (e.g., \textit{Rbot}); 
    \item \textit{Other}, for remaining attacks (e.g., \textit{shellcode}, \textit{scanning}).
\end{itemize}
We remark that, in our implementation of XeNIDS, we \textit{always} use the specific attack classes, i.e., we do not `aggregate' multiple attacks into a single class. The differentiation provided in Table~\ref{tab:dataset2} is for comprehensiveness, because the amount of specific attacks of our testbed is very broad.
As an example, \dataset{NB15} (and, hence, \dataset{UF-NB15}) has samples for all families, i.e., 2 different types of \textit{Botnet} attacks, 1 type of \textit{DoS}, and 6 types of \textit{Other} attacks; whereas \dataset{CTU13} only has samples for 5 different attacks of the \textit{Botnet} family. From Table~\ref{tab:dataset2} we also determine that $\mu$=46 in the \textit{heterogeneous} scenario, and that $\mu$=36 in the \textit{uniform} scenario---this is because all the specific attack types are distinct across the input $\mathbb{D}$ datasets.

\begin{table}[!ht]
    \centering
    \caption{Distribution of attacks in each dataset. In our implementation of XeNIDS, we \textit{always} use the specific attack classes, and perform no merging.}
    \resizebox{0.95\columnwidth}{!}{
        \begin{tabular}{cccc||cccc}
        \toprule
        \multicolumn{4}{c||}{\normalsize{\textit{Heterogeneous} scenario}} & \multicolumn{4}{c}{\normalsize{\textit{Uniform} scenario}} \\
        \textbf{Dataset} & \textbf{\textit{Botnet}} & \textbf{\textit{DoS}} & \textbf{\textit{Other}} & \textbf{Dataset} & \textbf{\textit{Botnet}} & \textbf{\textit{DoS}} & \textbf{\textit{Other}}\\ 
        
        \midrule
        \dataset{CTU13} & 5 & 0 & 0 & \dataset{UF-BotIoT} & 0 & 2 & 2\\
        \dataset{NB15} & 2 & 1 & 6 & \dataset{UF-NB15} & 2 & 1 & 6\\
        \dataset{IDS18} & 1 & 5 & 8 & \dataset{UF-IDS18} & 1 & 5 & 8\\
        \dataset{DDOS19} & 0 & 18 & 0 &  \dataset{UF-ToNIoT} & 0  & 2 & 7\\
        \midrule
        Total & 8 & 24 & 14 & Total & 3 & 10 & 23\\ 
        \bottomrule
        
        \end{tabular}
    }
    
    \label{tab:dataset2}
\end{table}

\subsection{Preprocessing}
\label{ssec:preprocessing}
We now describe the preprocessing computed on each considered NID dataset $D_i \in \mathbb{D}$ for both scenarios. Such operations represent the first two stages of XeNIDS: standardize (§\ref{ssec:standardize}) and isolate (§\ref{ssec:isolate}). Our low-level implementation of XeNIDS aims to overcome all the residual challenges in §\ref{ssec:challenges}---to the extent this is possible with the current state-of-the-art.
The experimental platform is an Ubuntu 20.04 machine with 64GB RAM and an Intel Xeon E5-2620 CPU. The development leverages the Scikit-Learn suite. 

\textbf{Standardize.} 
We first associate each sample to its ground truth\footnote{We verify the checksum of each dataset, if provided.}. Then, we derive a common feature set based on the official NetFlow v9 documentation, which we report in Table~\ref{tab:features}.
These features represent the minimum set of common features obtainable from the source data for both scenarios. We note that some datasets are provided with more features (e.g., \dataset{IDS18} has 80), which are left out. However, as we will show in our experiments, the considered features yield ML-NIDS with state-of-the-art performance.

\begin{table}[!h]
    \scriptsize
    \centering
    \caption{Feature set of our XeNIDS implementation in both scenarios.}
    \begin{tabular}{ccc}\toprule
    \textbf{\#} & \textbf{Feature Name} & \textbf{Type}\\ \midrule
    1 & Source IP address internal  & Bool \\
    2 & Destination IP address internal & Bool\\
    3 & Source port type & Cat\\
    4 & Destination port type & Cat\\
    5 & Flow Duration $[s]$ & Num \\
    6 & Flow Direction & Bool\\
    7 & Incoming Bytes & Num\\
    8 & Outgoing Bytes & Num\\
    9 & Total Bytes& Num\\
    10 & Incoming Packets & Num\\
    11 & Outgoing Packets & Num\\
    12 & Total Packets & Num\\
    
    \bottomrule
    
    \end{tabular}
    
    \label{tab:features}
\end{table}
To sanitize network artifacts, we follow the recommendations in Appendix~\ref{app:recommendations}.
To avoid overfitting and simulate the application of anonymisation techniques, we do not use the plain IP addresses or service-ports as features. Instead, we differentiate between internal/external hosts of each network (features 1 and 2 in Table~\ref{tab:features}); and we categorize the network ports according to the IANA guidelines (features 3 and 4 in Table~\ref{tab:features}). All of these operations are also adopted by recent works (e.g.~\cite{apruzzese2020deep}). We set the $d$ of all samples in seconds, and we ensure that most samples fall within the same duration range (i.e., [0-150]s), discarding the few outliers.

\vspace{1mm}

\textbf{Isolate.}
For each dataset $D_i$ in $\mathbb{D}$, we separate benign from malicious samples using the ground truth label. We do not make any aggregation, hence our $\mu$ are the original ones (i.e., $\mu$=46 for the \textit{heterogeneous} scenario, and $\mu$=36 for the \textit{uniform} scenario). We thus obtain the following:
\begin{itemize}
    \item for the \textit{heterogeneous} scenario, $\mathbb{N}$ containing 4 elements representing the source networks of the respective datasets (\dataset{CTU13}, \dataset{NB15}, \dataset{IDS18}, \dataset{DDOS19}), and $\mathbb{M}$ containing 184 elements (because $n$=4 and $\mu$=46);
    \item for the \textit{uniform} scenario, $\mathbb{N}$ containing 4 elements representing the source networks of the respective datasets (\dataset{UF-BotIoT}, \dataset{UF-NB15}, \dataset{UF-IDS18}, \dataset{UF-TonIoT}), and $\mathbb{M}$ containing 144 elements (because $n$=4 and $\mu$=36).
\end{itemize}
XeNIDS can now create the contexts to be cross-evaluated.

\subsection{Assessment}
\label{ssec:assessment}
In both scenarios we analyse three context types: \C{1}, \C{4} and \C{7} (cf. Table~\ref{tab:contexts}). Let us explain the common assessment procedures, focusing on the architecture of the ML-NIDS and the considered performance metrics.

\textbf{Parameters and Performance Metrics.}
We use the same parameters for our implementation of XeNIDS. Specifically, the adopted splits $s(N)$ and $s(M)$ are always 80:20 for both $T$ and $E$. We use such splits because they are common in related literature (e.g.,~\cite{apruzzese2020deep, zhang2021network}), therefore enabling a more fair comparison of our results with those of past works.
We considered different ML algorithms, but we found that Random Forests consistently provided the best tradeoff in terms of detection performance, rate of false alarms, and training time---a result that confirms the state-of-the-art on the same datasets (e.g.,~\cite{zhang2021network, apruzzese2020deep, pontes2021potts, sarhan2020netflow}). Hence our results will refer to Random Forest as the learning algorithm for each classifier.  
The performance metrics of interest are the \textit{F1-score} (F1) and the \textit{False Positive Rate} (FPR), defined as follows:
\begin{tabularx}{\columnwidth}{XX}
    \begin{equation*}
        \label{exp:fscore}
        \small{\text{F1}\!=\! \frac{tp}{.5(fp\!+\!fn)\!+\!tp},}
    \end{equation*}
    &
    \begin{equation}
        \label{exp:fpr}
        \small{\text{FPR}\!=\!\frac{fp}{tp\!+\!fn},}
    \end{equation}
\end{tabularx}
\noindent
where $tp$, $fp$, $fn$ denote true positives, false positives, and false negatives, respectively; we consider a ``true positive'' as the correct detection of a malicious sample. It is desirable that the application of XeNIDS when considering modifications of the training set (hence, \C{7}) should preserve the baseline FPR (cf. §\ref{ssec:challenges}). Finally, to account for the randomness of each split, we repeat each experiment 5 times, and in our results, we will report the average of each repetition.

\textbf{ML-NIDS Architecture.}
XeNIDS fosters development of \textit{ensembles} of detectors (§\ref{ssec:contextualize}), each specialized in a single attack. There are many ways in which such detectors can be integrated in a ML-NIDS.
In our implementation, we assume that the NIDS uses ML as a final confirmation of detection. Hence, each sample is forwarded to the `most suitable' detector of the ensemble, which must determine whether such sample is really malicious or not. While such selection is straightforward for contexts of type \C{1} and \C{7} (because 
$\bar{\tau}=\bar{\varepsilon}$), this is not the case for \C{4}, where the ML-NIDS is tested against `unknown' attacks (because $\bar{\tau}\neq\bar{\varepsilon}$). Hence, for \C{4} we perform a preliminary exploratory operation to identify the most suitable detector of the ensemble against the unknown attacks; this is to allow a more fair comparison with~\cite{pontes2021potts}, which also investigates \C{4} by using two datasets considered in our testbed, \dataset{IDS18} and \dataset{DDOS19}.
Hence, we reserve a portion of the samples of each unknown malicious class, and test every detector composing the ML-NIDS against such portion: the one with the best performance is chosen as the candidate for analyzing the corresponding attack. This is legitimate because the ground truth of such samples is known, and such samples (i) are never used in $E$, and (ii) are not added in $T$ (otherwise it would not be \C{4}). Therefore, when presenting the corresponding results, we will report the performance achieved by the most optimal detector against each specific attack.

%% file: sections/6-demonstration.tex
\section{Demonstration}
\label{sec:demonstration}
Our demonstration aims to simulate the exemplary use-cases described in §\ref{sssec:usecase}. Let us discuss how we organize our demonstration by using the three considered types of contexts (\C{1}, \C{4} and \C{7}) enabled by the proposed model (cf. Table~\ref{tab:contexts}). %

\textbf{Workflow.} We follow the same workflow for both the \textit{uniform} and \textit{heterogeneous} scenario. 
\begin{enumerate}
    \item \RQ{Baseline} (§\ref{ssec:baseline}). We begin by assessing the case where the organization $\mathcal{O}$ has an $N$ and a $M$ collected in their own network $o$. Such setup corresponds to context \C{1}. To simulate \C{1}, we use XeNIDS to devise a ML-NIDS for each dataset; such ML-NIDS is composed by an ensemble of detectors, each trained on a single attack contained in the same dataset. The ML-NIDS is tested against all the attacks of the `origin' dataset. The expectation is that the results match the state-of-the-art.
    \item \RQ{Generalization} §\ref{ssec:generalization}. Having a ML-NIDS, the organization $\mathcal{O}$ wants to assess its effectiveness against different attacks not included in $T$ and originating from a different network than $o$. Such setup corresponds to context \C{4}. We use XeNIDS to \textit{test} the `baseline' detectors of \C{1} against \textit{all} attacks of \textit{all} datasets. The expectation is that the performance will decrease substantially.
    \item \RQ{Extension} (§\ref{ssec:extension}). To compensate for the low performance against unknown attacks, the organization $\mathcal{O}$ borrows more malicious samples to improve the detection capabilities of their ML-NIDS. This corresponds to context \C{7} where \smallset{t} extends \smallset{o}. For each dataset, XeNIDS trains additional detectors by using the malicious samples of \textit{all} the other datasets, and adds such detectors to the ensemble of the `baseline' ML-NIDS. Such `extended' ML-NIDS is tested against \textit{all} attacks of the CS. The expected result is an improved performance w.r.t. \C{4}. 
    \item \RQ{Surrogation} (§\ref{ssec:surrogation}). If the organization $\mathcal{O}$ only has benign samples $N$ from their own network $o$ but does not have an $M$, the only option is using an $M$ from a different network to develop a `surrogate' ML-NIDS. This is also represented by \C{7}, but in this case \smallarray{o} and \smallarray{t} are disjointed. Hence, for each dataset, XeNIDS uses \textit{only} the additional detectors developed at the previous step to devise a (new) `surrogate' ML-NIDS. Such surrogate ML-NIDS is then tested only against the `attacks from different networks.
\end{enumerate}

\textbf{Example.} Let us provide a complete example of the workflow above. We adopt the viewpoint of an organization that owns the \dataset{UF-UNB15} network (hence, the \textit{uniform} scenario). A total of 9 attacks originate from such network: 2 botnets, 1 DoS, and 6 others (cf. Table~\ref{tab:dataset2}). 
\begin{enumerate}
    \item XeNIDS uses the 9 attacks of \dataset{UF-UNB15} to train 9 detectors, representing the baseline ML-NIDS, which is tested against these 9 attacks.
    \item XeNIDS tests the baseline ML-NIDS (with its 9 detectors) against all the attacks of all datasets. Namely: 4 attacks for \dataset{UF-BotIoT}, 14 attacks for \dataset{UF-IDS18}, 9 attacks for \dataset{UF-ToNIoT}, as well as the 9 in \dataset{UF-UNB15}.
    \item XeNIDS trains 27 additional detectors, each using the benign samples of \dataset{UF-UNB15} alongside the malicious samples of a specific attack contained in \dataset{UF-IDS18}, \dataset{UF-ToNIoT}, \dataset{UF-BotIoT}, respectively. Such detectors are \textit{combined} with the 9 `baseline' detectors of \dataset{UF-UNB15}, to extend the ML-NIDS. Such `extended' ML-NIDS is tested against \textit{all} the attacks of \textit{all} datasets (36 attacks). 
    \item XeNIDS uses only the 27 detectors trained in the previous step (representing the `surrogate' ML-NIDS) and tests them against the attacks contained in the corresponding networks, i.e., without taking into account the attacks (and the detectors) in \dataset{UF-UNB15}.
\end{enumerate}
Such workflow is followed 4 times for both scenarios, each time by considering a different dataset as `origin'.

\input{sections/6.1-baseline}

\input{sections/6.2-generalization}
\input{sections/6.3-extension}
\input{sections/6.4-surrogation}

%% file: sections/6.1-baseline.tex
\subsection{\RQ{Baseline}}
\label{ssec:baseline}
We start by assessing \C{1}, and report the results in Table~\ref{tab:baseline}.
Specifically, on the left, we present the results for the \textit{heterogeneous} scenario, and on the right, the \textit{uniform} scenario.
For each dataset, we report the average F1-score obtained against each family of attacks (cf. Table~\ref{tab:dataset2}).
Moreover, we report in the captions the average \textit{FPR} achieved by the ML-NIDS of each dataset. Henceforth, all our results will be reported in the same format as Table~\ref{tab:baseline}.

\begin{table}[!ht]
    \centering
    \caption{\RQ{Baseline}--\captionC{1}. The table shows the F1-score of the ML-NIDS for each origin network. The performance matches the state-of-the-art. The FPR is less than 0.001 for all networks aside from \captiondataset{UF-BoTIoT} (FPR = 0.11).}
    \resizebox{0.95\columnwidth}{!}{
        \begin{tabular}{c|ccc||c|ccc}
        \toprule
        \multicolumn{4}{c||}{\normalsize{\textit{Heterogeneous} scenario}} & \multicolumn{4}{c}{\normalsize{\textit{Uniform} scenario}} \\
        \textbf{Dataset} & \textbf{\textit{Botnet}} & \textbf{\textit{DoS}} & \textbf{\textit{Other}} & \textbf{Dataset} & \textbf{\textit{Botnet}} & \textbf{\textit{DoS}} & \textbf{\textit{Other}}\\ 
        
        \midrule
        \captiondataset{CTU13} & 98.1 & --- & --- & \captiondataset{UF-BotIoT} & --- & 99.9 & 92.0 \\
        \captiondataset{NB15} & 88.6  & 98.5 & 98.1  & \captiondataset{UF-NB15} & 83.4 & 91.4 & 95.8\\
        \captiondataset{IDS18} & 90.1 & 99.9  & 96.1 & \captiondataset{UF-IDS18} & 99.9 & 99.1 & 99.2\\
        \captiondataset{DDOS19} & ---  & 99.9  & ---  &  \captiondataset{UF-ToNIoT} & --- & 99.9 & 99.7\\
        \bottomrule
        \end{tabular}
    }
    \label{tab:baseline}
\end{table}

From Table~\ref{tab:baseline}, we observe that there are only two scores for \dataset{UF-BoTIoT} and \dataset{UF-ToNIoT}, because there are no \textit{Botnet} samples in these `origin' datasets. Similarly, \dataset{CTU13} and \dataset{DDOS19} presents only one score.

All our \RQ{baseline} detectors match the performance of past works (cf. Table~\ref{tab:baseline}). As an example, for the \textit{uniform} scenario, the `worst' ML-NIDS is trained (and evaluated) on \dataset{UF-UNB15}, but also in~\cite{sarhan2020netflow} such ML-NIDS achieves an average F1-score of $85\%$. Similarly, in the \textit{heterogeneous} scenario, the ML-NIDS in~\cite{pontes2021potts} achieve 99.0 F1-score on both \dataset{IDS18} and \dataset{DDOS19}, whereas~\cite{apruzzese2020deep} achieves 99.0 F1 on \dataset{CTU13}---all these results align with ours, confirming that our XeNIDS implementation is efficient.

%% file: sections/6.2-generalization.tex
\subsection{\RQ{Generalization}}
\label{ssec:generalization}

We then assess the baseline ML-NIDS when they are subject to attacks also contained in different networks, i.e., \C{4}. We report the detection results in Table~\ref{tab:generalization}; the FPR is the same as in the baseline \C{1} (cf. caption of Table~\ref{tab:baseline}): this is expected because in \C{4} uses the same training sets as \C{1}, and also the benign samples of the evaluation sets are the same as in \C{1},

From Table~\ref{tab:generalization}, we observe that the performance decreases because most of the attacks are `unknown' to the \RQ{baseline} ML-NIDS.
However, we can observe some interesting phenomena. 

In the \textit{uniform} scenario, the ML-NIDS of \dataset{UF-ToNIoT} can detect botnet attacks remarkably well ($82\%$ F1-score), despite \textit{having no} ML-model specialized on botnet attacks (because no such attacks are contained in \dataset{UF-ToNIoT}). 
Such an intriguing finding could only be appreciated by cross-evaluating the ML-NIDS trained on \dataset{UF-ToNIoT} against malicious samples from different networks.
Furthermore, the \textit{heterogeneous} scenario shows that the baseline ML-NIDS of \dataset{DDOS19} works very well against DoS attacks from other networks---despite such attacks being performed by different means. 

We also compare some of our results with those in~\cite{pontes2021potts}, which also investigated \C{4}. 
Specifically, the ML-NIDS trained on \dataset{DDOS19} and tested on \dataset{IDS18} in~\cite{pontes2021potts} achieves $64\%$ F1-score on average, which is similar to ours. 
Conversely, the ML-NIDS trained on \dataset{IDS18} and tested on \dataset{DDOS19} in~\cite{pontes2021potts} achieves an average $78\%$ F1-score, which is slightly superior than ours. We explain this difference to the different conditions in~\cite{pontes2021potts}: they only consider a smaller portion of the initial dataset, whereas we use all of them. Hence, our samples may present a more skewed distribution that makes them more difficult to classify.

\begin{table}[!ht]
    \centering
    \caption{\RQ{Generalization}--\captionC{4}. Each baseline ML-NIDS of \captionC{1} is tested against the attacks of all other networks. Most attacks are not detected, and the F1-score degrades. The FPR is the same as in \captionC{1} because the benign samples are always the same and the training set is not modified.}
    \resizebox{0.95\columnwidth}{!}{
        \begin{tabular}{c|ccc||c|ccc}
        \toprule
        \multicolumn{4}{c||}{\normalsize{\textit{Heterogeneous} scenario}} & \multicolumn{4}{c}{\normalsize{\textit{Uniform} scenario}} \\
        \textbf{Dataset} & \textbf{\textit{Botnet}} & \textbf{\textit{DoS}} & \textbf{\textit{Other}} & \textbf{Dataset} & \textbf{\textit{Botnet}} & \textbf{\textit{DoS}} & \textbf{\textit{Other}}\\ 
        
        \midrule
        \captiondataset{CTU13} & 80.0  & 38.1 & 49.7 & \captiondataset{UF-BotIoT} & 47.8 & 69.0 & 76.8 \\
        \captiondataset{NB15} & 65.8 & 40.7 & 75.2 & \captiondataset{UF-NB15} & 72.2 & 52.3 & 64.1\\
        \captiondataset{IDS18} & 54.9 & 49.4 & 76.1 & \captiondataset{UF-IDS18} & 68.2 & 81.0 & 63.3\\
        \captiondataset{DDOS19} & 54.4 & 99.5 & 83.1 &  \captiondataset{UF-ToNIoT} &  82.1 & 89.3 & 85.1\\
        \bottomrule
        \end{tabular}
    }
    \label{tab:generalization}
\end{table}

%% file: sections/6.3-extension.tex
\subsection{\RQ{Extension}}
\label{ssec:extension}

Next, we assess \C{7} when $\bar{\tau}$ extends $\bar{o}$, and report the results in Table~\ref{tab:extension}. 
We observe that the overall performance increases (w.r.t. Table~\ref{tab:generalization}) by augmenting the training sets with the corresponding malicious samples. 

In the heterogeneous scenario, our `extended' ML-NIDS naturally outperform those in~\cite{pontes2021potts}, but we cannot claim this as a contribution because our `extended' ML-NIDS use an augmented training set.

We also appreciate that the FPR remains stable (cf. Table~\ref{tab:baseline}). We owe such results to our reliance on ensembles of detectors.

\begin{table}[!ht]
    \centering
    \caption{\RQ{Extension}--\captionC{7}. By augmenting the training set of the ML-NIDS with the malicious samples, the F1-score improves w.r.t. \captionC{4}. The average FPR is lower than 0.001 for all networks aside from \captiondataset{UF-BotIoT} (FPR = 0.01).
    }
    \resizebox{0.95\columnwidth}{!}{
        \begin{tabular}{c|ccc||c|ccc}
        \toprule
        \multicolumn{4}{c||}{\normalsize{\textit{Heterogeneous} scenario}} & \multicolumn{4}{c}{\normalsize{\textit{Uniform} scenario}} \\
        \textbf{Dataset} & \textbf{\textit{Botnet}} & \textbf{\textit{DoS}} & \textbf{\textit{Other}} & \textbf{Dataset} & \textbf{\textit{Botnet}} & \textbf{\textit{DoS}} & \textbf{\textit{Other}}\\ 
        
        \midrule
        \captiondataset{CTU13} & 98.8  & 99.9 & 98.9 & \captiondataset{UF-BotIoT} & 99.7  & 99.9 & 99.2\\
        \captiondataset{NB15} & 97.1 & 99.9 & 99.1 & \captiondataset{UF-NB15} & 88.9 & 99.2 & 98.7\\
        \captiondataset{IDS18} & 98.5 & 99.7 & 97.7 & \captiondataset{UF-IDS18} & 99.9 & 99.4 & 97.8\\
        \captiondataset{DDOS19} & 99.9 & 99.9 & 98.6 &  \captiondataset{UF-ToNIoT} & 99.7  & 99.9 & 99.9\\
        \bottomrule
        \end{tabular}
    }
    
    \label{tab:extension}
\end{table}

%% file: sections/6.4-surrogation.tex
\subsection{\RQ{Surrogation}}
\label{ssec:surrogation}
Finally, we assess \C{7} when $\bar{\tau}$ and $\bar{o}$ are disjointed, and report the results in Table~\ref{tab:surrogation}. 
From this table, we observe that all detectors exhibit very high F1-scores, implying that the malicious samples are considerably different than the benign samples. Sometimes, the F1-score reaches $99.9\%$, but is not perfect: we believe such occurrence to be \textit{positive} because an F1-score of $100\%$ could be related to overfitting.

\begin{table}[!ht]
    \centering
    \caption{\RQ{Surrogation}--\captionC{7}. We exclude all malicious samples (and detectors) from the each `origin' network. The extremely high performance must be investigated. The average FPR is less than 0.001 and 0.0001 for all networks of \textit{uniform} and \textit{heterogeneous} scenarios, respectively.}
    \resizebox{0.95\columnwidth}{!}{
        \begin{tabular}{c|ccc||c|ccc}
        \toprule
        \multicolumn{4}{c||}{\normalsize{\textit{Heterogeneous} scenario}} & \multicolumn{4}{c}{\normalsize{\textit{Uniform} scenario}} \\
        \textbf{Dataset} & \textbf{\textit{Botnet}} & \textbf{\textit{DoS}} & \textbf{\textit{Other}} & \textbf{Dataset} & \textbf{\textit{Botnet}} & \textbf{\textit{DoS}} & \textbf{\textit{Other}}\\ 
        
        \midrule
        \captiondataset{CTU13} & 99.9  & 99.9 & 98.9 & \captiondataset{UF-BotIoT} & 99.7 & 99.9 & 99.9 \\
        \captiondataset{NB15} & 99.9 & 99.9 & 99.9 & \captiondataset{UF-NB15} & 99.9 & 99.9 & 99.9\\
        \captiondataset{IDS18} & 99.6 & 99.6 & 99.8 & \captiondataset{UF-IDS18} & 99.9  & 99.9 & 99.9\\
        \captiondataset{DDOS19} & 99.9 & 99.9 & 98.6 & \captiondataset{UF-ToNIoT} & 99.7 & 99.9 & 99.9\\
        \bottomrule
        \end{tabular}
    }
    
    \label{tab:surrogation}
\end{table}

%% file: sections/7-discussion.tex
\section{Discussion}
\label{sec:discussion}
We now discuss the results presented in §\ref{sec:demonstration}. We first summarize the main findings (§\ref{ssec:analysis}), and then make some considerations \textit{reliability} of the results (§\ref{ssec:analysis_uf} and §\ref{ssec:analysis_het}). We then present the main limitations of our demonstration, as well as possible workarounds (§\ref{ssec:limitations}).

\subsection{Preliminary Analysis}
\label{ssec:analysis}
We appreciate that, in general, our results show the effectiveness of XeNIDS in producing baselines with state-of-the-art performance, while also extending the detection surface. It is intriguing that, in some cases, it is possible to detect attacks without training on the related malicious samples. 
To further stress the advantages of cross-evaluations, we provide an in-depth look at our results by focusing on the \dataset{CTU13} dataset. This dataset contains \textit{only} \textit{botnet} attacks and, from Table~\ref{tab:dataset}, the state-of-the-art (e.g.,~\cite{apruzzese2020deep}) achieves $99.1$\% F1-score against such attacks. A similar performance may suggest that improvements can be incremental at best; however, no past works have assessed how ML-NIDS trained on \dataset{CTU13} can detect different \textit{botnet} attacks (not included in \dataset{CTU13}). By applying the proposed XeNIDS framework, we discover that similar ML-NIDS perform much worse: as shown by Table~\ref{tab:generalization} (§\ref{ssec:generalization}), the F1-score of such ML-NIDS drops by 20\% against \textit{botnet} attacks of diverse datasets; even worse, it is unable to detect \textit{DoS} attacks (F1-score of $38$\%). \textit{Such poor performance could only be assessed via cross-evaluations.} To make it better, the performance against these -- different -- attacks can be increased by training on the respective samples: by observing Table~\ref{tab:extension}, the F1-score can be restored to 99\% via cross-evaluations. Also noteworthy is that the FPR always remains within acceptable levels (below 0.001). Such FPR will resemble the one after deployment (because the source of benign samples is always the same).

However, as stated in §\ref{ssec:challenges}, it is necessary to further analyze the results of XeNIDS. This is to avoid relying on a false-sense of security, given by high performance at test-time which does not correspond to the performance after the ML-NIDS is deployed. We specifically focus on contexts of type \C{7} because they involve modifications of the training data, which can lead to `network artifacts' that affects the $Env$ component of NetFlows (cf. Exp.~\ref{exp:env} in §\ref{ssec:standardize}) and, potentially, lead to overfitted ML-NIDS.

\subsection{Reliability: Uniform scenario}
\label{ssec:analysis_uf}
In this scenario, by definition, the $Env$ is affected only by $NetId$ because $Conf$ is the same for all datasets; such characteristic implicitly reduces the risk of network artifacts. Nevertheless, we find instructive to analyze the results of the \RQ{surrogate} ML-NIDS, reported in the right-side of Table~\ref{tab:surrogation}. In particular, we consider the \dataset{UF-UNB15} network. We observe that the `surrogate' detectors focused on botnet attacks achieve a near-perfect F1-score, which is \textit{higher} than both their `extended'  and `baseline' variants (cf. Tables~\ref{tab:extension} and~\ref{tab:baseline}). This implies that benign samples of \dataset{UF-UNB15} are very similar to the (malicious) botnet samples of \dataset{UF-UNB15}, making such botnet samples harder to classify by the \dataset{UF-UNB15} ML-NIDS w.r.t. the botnet samples in other networks. Such occurrence can be a sign of overfitting, because the \dataset{UF-UNB15} ML-NIDS could be detecting the botnet samples from other networks on the basis of network artifacts. However, a more detailed analysis can remove such doubt. Indeed, in the uniform scenario, the only other source of `botnet' samples is \dataset{UF-IDS18}, where the baseline performance is also perfect (cf. Table~\ref{tab:baseline}), a result also confirmed by the state-of-the-art~\cite{sarhan2020netflow}. Simply put, the `botnet' samples in \dataset{UF-IDS18} are easy to identify. Such observation reduces the chance that the \RQ{surrogate} (or the \RQ{extended}) ML-NIDS of \dataset{UF-UNB15} are affected by artifacts from \dataset{UF-IDS18}.

\subsection{Reliability: Heterogeneous scenario}
\label{ssec:analysis_het}
This scenario assumes NetFlows generated via different means, hence the $Env$ component is affected by both $NetId$ and $Conf$. Such characteristic increases the chance that some artifacts `evaded' XeNIDS standardize stage.
To find a trace of such artifacts, we compare the \textit{feature importances} of each ML-NIDS (\textit{all} ML-NIDS use the same feature set).

\begin{figure}[!htbp]
    \centering
    \includegraphics[width=0.85\linewidth]{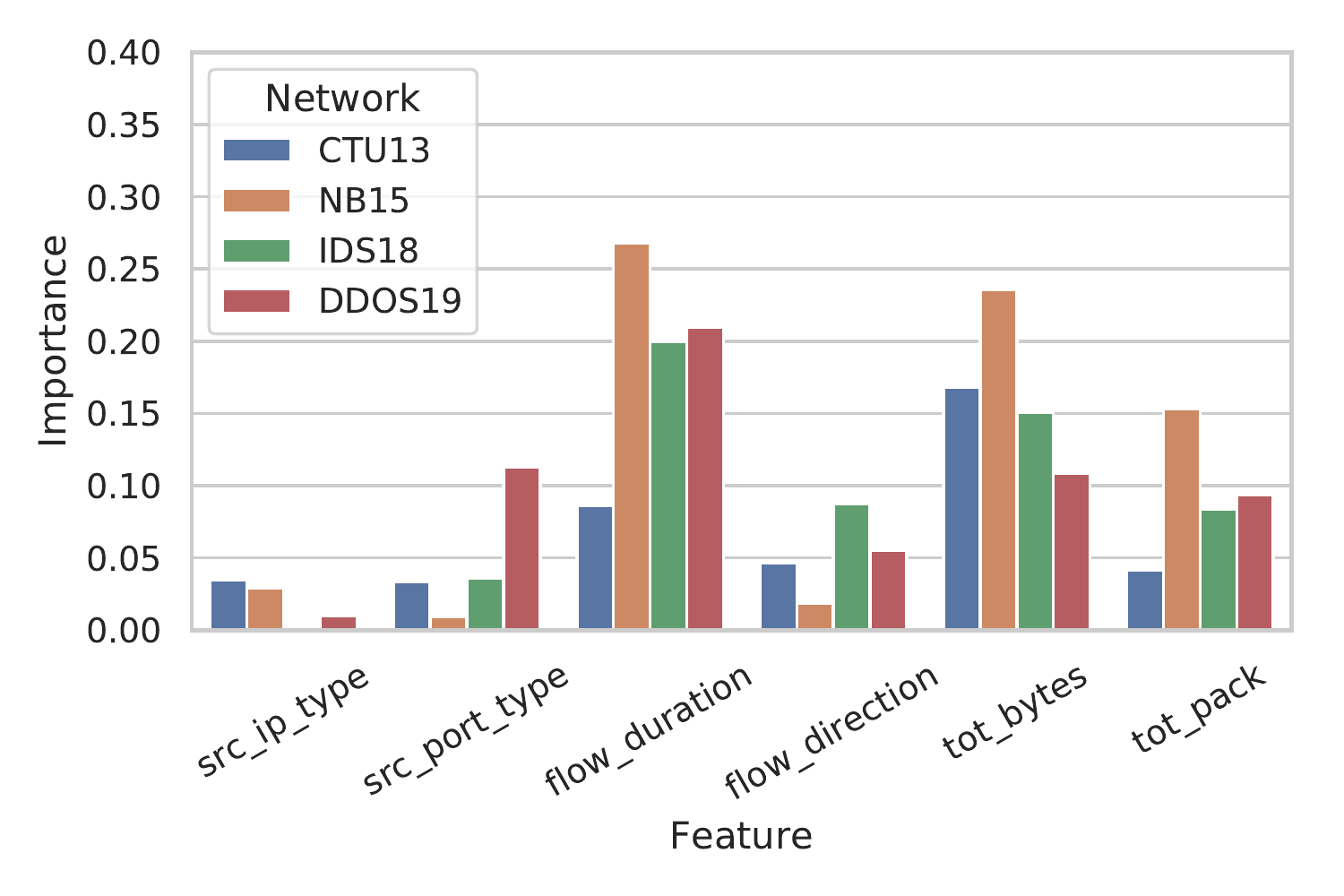}
    \caption{Feature importances of the \textit{Rbot} detectors (Heterogeneous CS).}
    \label{fig:imp.ctu13}
\end{figure}

Intuitively, the most important features for detecting an attack in its `origin' network should denote the malicious behavior--hence, such features should be also the most important when the attack is `transferred' to train a different ML-NIDS (which is the case in \C{7}).
We provide in Fig.~\ref{fig:imp.ctu13} a comparison of such importances, focusing on the detectors specialized on the \textit{Rbot} botnet attack (contained in the \dataset{CTU13} network). Specifically, Fig.~\ref{fig:imp.ctu13} shows the importances of the top6 most important features (out of 12--cf. Table~\ref{tab:features}) for all the \textit{Rbot} detectors among the four different networks. 

From Fig.~\ref{fig:imp.ctu13} we observe that the detectors can either `agree' or `disagree' on the importance of such features. Specifically, we observe that the `origin' \dataset{CTU13} detector (blue bars) places a great importance on the \textit{tot\_bytes}, denoting agreement with the other detectors; however, there is disagreement on the \textit{duration}, which is less important for the \dataset{CTU13} detector. The general trend in Fig.~\ref{fig:imp.ctu13} is that the detectors disagree on most features: therefore, we cannot exclude that some underlying effects of $Env$ are still present.

\subsection{Limitations and Future Work}
\label{ssec:limitations}
To increase the reliability of the detection performance, it is necessary to assume the perspective of the \textit{owners} of each network. As a practical example that could remove any doubt, the owners of the \dataset{NB15} network should infect their machines with the \textit{Rbot} botnet (contained in \dataset{CTU13}), and verify whether their ML-NIDS (trained on the \textit{Rbot} samples from \dataset{CTU13}) can detect such attack. Doing such verifications is not possible for our scientific paper, as they require a complete control and overview of the monitored network. Moreover, the CnC servers of the \textit{Rbot} botnet are no longer active. Our experiments are for demonstrative purposes, but realistic deployments should integrate such verifications--which \textit{must} be done regardless of the origin of the malicious samples (i.e., both in `traditional' and in `cross' evaluations).

Moreover, we note that each considered context type is an independent use case. Indeed, our focus is not on developing systems that outperform the state-of-the-art: it would be unfair to claim that our ML-NIDS generated via \C{7} are better than those in~\cite{pontes2021potts}. In contrast, our goal is to demonstrate the contexts that can be assessed by mixing different network datasets, showcasing the potential of such cross-evaluation for the state-of-the-art. As such, we can consider our results as a `benchmark', allowing future cross-evaluation studies to compare their results with those in our paper. 

Finally, an intriguing future research direction is the assessment of cross-evaluations in adversarial settings: for instance, how would a ML-NIDS poisoned with samples from a different network perform (cf. §\ref{ssec:challenges})? Answering a similar question would be beneficial to the ML-NIDS research area.

%% file: sections/8-conclusions.tex
\section{Conclusions}
\label{sec:7-conclusions}

Despite many successes, the integration of supervised Machine Learning (ML) methods in Network Intrusion Detection Systems (NIDS) is still at an early stage. This is due to the difficulty in obtaining comprehensive sets of labelled data for training and evaluating a ML-NIDS. The recent release of labelled datasets for ML-NIDS was appreciated by the research community; however, few works noticed the opportunity that such availability provides to the state-of-the-art.

Inspired by the necessity of proactive empirical evaluations and the recent release of more open datasets, we promote the idea of cross-evaluating ML-NIDS by using existing labelled data from different networks. Such approach has been applied before, but no past work specifically tackled this problem. As a result, all the benefits of cross-evaluations, as well as their intrinsic risks, are still unexplored.

We address all of these issues in this paper. We begin by presenting the first model for cross-evaluation of ML-NIDS, which is data-agnostic and general enough to cover both supervised and unsupervised ML-NIDS. By using such model, we highlight the limited scope adopted by most related works, and showcase the benefits provided by cross-evaluations of ML-NIDS. We also present all the challenges and limitations of such opportunity, which must be known and adequately addressed in order to provide actionable results.

To foster proactive cross-evaluations, we develop XeNIDS, the first framework for cross-evaluations of ML-NIDS. XeNIDS aims to mitigate all the hazards arising from using data from different networks. Specifically, XeNIDS focuses on NetFlow data, which is popular in the ML-NIDS community due to its flexibility and suitability for detection purposes. 

Finally, we elucidate the potential of cross-evaluations via a large set of experiments, where we use XeNIDS to cross-evaluate ML-NIDS on 6 well-known datasets. In our demonstration, we show the capability of XeNIDS to retain the `baseline' performance of past ML-NIDS, while illustrating some additional use-cases enabled by cross-evaluations, such as `extending' the detection surface of ML-NIDS. We conclude our demonstration with a follow-up discussion where we question the reliability of the results, which is necessary for realistic deployments of ML-NIDS.

Our paper will hopefully inspire future works on ML-NIDS, and is oriented to both researchers and practitioners. The former can make better use of open datasets to cross-evaluate past and future ML-NIDS, allowing broader assessments of the state-of-the-art; the latter can use future research results, or completely integrate cross-evaluations in their proactive assessments, to develop or improve Machine Learning-based Network Intrusion Detection Systems---without incurring in extra labelling procedures. We believe that cross-evaluations -- supported by data-sharing platforms and federated learning techniques -- represent a pragmatic way to overcome the specificity of NIDS and realize `general' ML-NIDS.

%% file: bibliography.tex

%% file: biographies/bio.tex
\begin{IEEEbiography}
  [{\includegraphics[width=1in,height=1.25in,keepaspectratio]{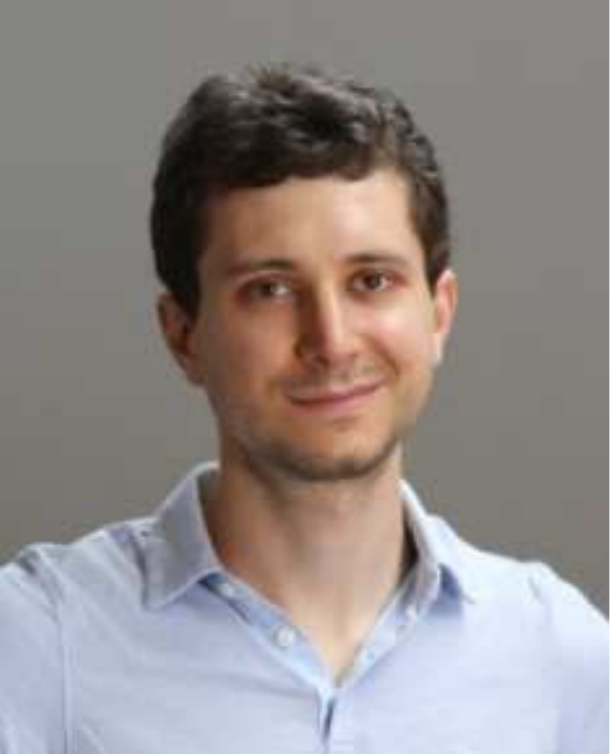}}]{Giovanni Apruzzese} is a Post-Doctoral researcher within the Institute of Information Systems at the University of Liechtenstein since 2020. He received the PhD Degree and the Master's Degree in Computer Engineering (summa cum laude) in 2020 and 2016 respectively at the University of Modena, Italy. In 2019 he spent 6 months as a Visiting Researcher at Dartmouth College (Hanover, NH, USA) under the supervision of Prof. VS Subrahmanian. His research interests involve all aspects of big data security analytics with a focus on machine learning, and his main expertise lies in the analysis of Network Intrusions, Phishing, and Adversarial Attacks.
  
  Homepage: \url{https://www.uni.li/giovanni.apruzzese}
\end{IEEEbiography}
\begin{IEEEbiography}
 [{\includegraphics[width=1in,height=1.25in,clip,keepaspectratio]{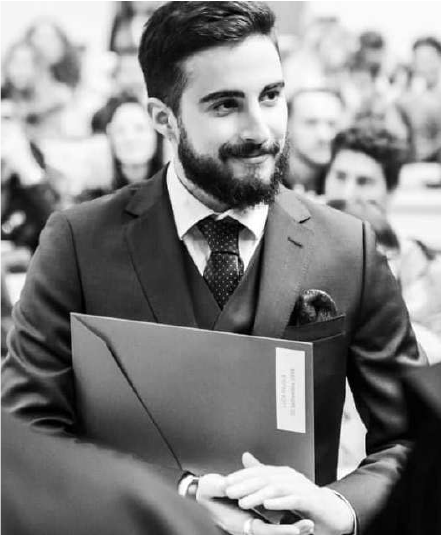}}]{Luca Pajola} is currently a Ph.D. student in school of Brain Mind and Computer Science at the University of Padova, Italy. Here, he is part of the SPRITZ Security and Privacy Research Group research group under the supervision of Prof. Mauro Conti. He received my MSc in Computer Science in 2018 at University of Padova, Italy. He is conducting research on fields including security and machine learning.
Homepage: \url{https://www.math.unipd.it/~pajola/}
\end{IEEEbiography}
\begin{IEEEbiography}
 [{\includegraphics[width=1in,height=1.25in,clip,keepaspectratio]{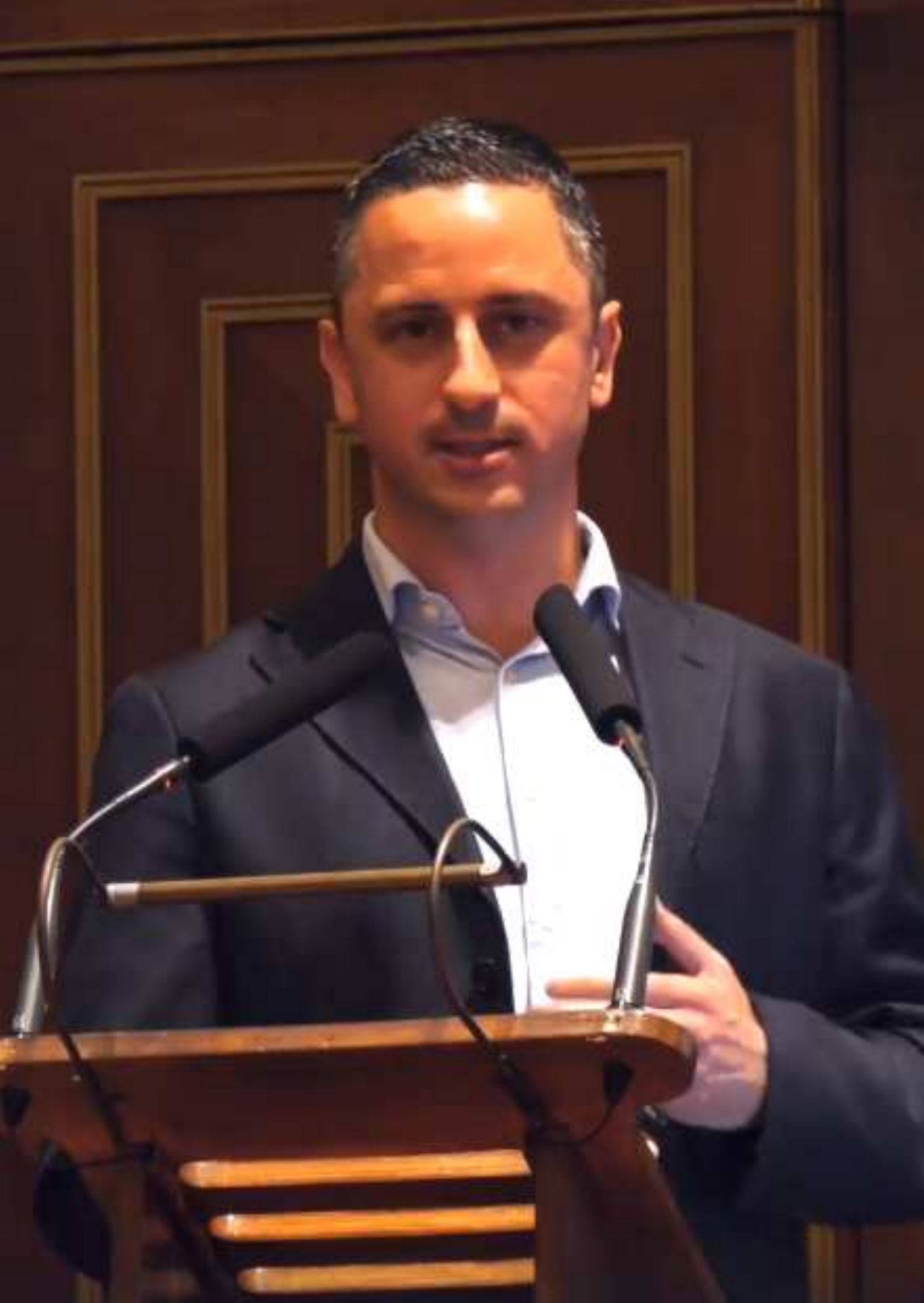}}]{Mauro Conti} Mauro Conti is Full Professor at the University of Padua, Italy. He is also affiliated with TU Delft and University of Washington, Seattle. He obtained his Ph.D. from Sapienza University of Rome, Italy, in 2009. After his Ph.D., he was a Post-Doc Researcher at Vrije Universiteit Amsterdam, The Netherlands. In 2011 he joined as Assistant Professor at the University of Padua, where he became Associate Professor in 2015, and Full Professor in 2018. He has been Visiting Researcher at GMU, UCLA, UCI, TU Darmstadt, UF, and FIU. He has been awarded with a Marie Curie Fellowship (2012) by the European Commission, and with a Fellowship by the German DAAD (2013). His research is also funded by companies, including Cisco, Intel, and Huawei. His main research interest is in the area of Security and Privacy. In this area, he published more than 400 papers in topmost international peer-reviewed journals and conferences. He is Editor-in-Chief for IEEE Transactions on Information Forensics and Security, Area Editor-in-Chief for IEEE Communications Surveys \& Tutorials, and has been Associate Editor for several journals, including IEEE Communications Surveys \& Tutorials, IEEE Transactions on Dependable and Secure Computing, IEEE Transactions on Information Forensics and Security, and IEEE Transactions on Network and Service Management. He was Program Chair for TRUST 2015, ICISS 2016, WiSec 2017, ACNS 2020, CANS 2021, and General Chair for SecureComm 2012, SACMAT 2013, NSS 2021 and ACNS 2022. He is Fellow of the IEEE, Senior Member of the ACM, and Fellow of the Young Academy of Europe.
 Homepage: \url{https://www.math.unipd.it/~conti/}
\end{IEEEbiography}

%% file: sections/appendix.tex
\section{Contributors to NetFlows}
\label{app:influence}
Let us illustrate the role played by $Comm$ and $Env$ (i.e., $NetId$ and $Conf$) in the generation of the corresponding NetFlows (see Exp.~\ref{exp:env}).
Assume that two organizations, $\mathcal{O}_1$ and $\mathcal{O}_2$, have two distinct networks both having a pair of hosts ($h_1^1$ and $h_1^2$ for $\mathcal{O}_1$, $h_2^1$ and $h_2^2$ for $\mathcal{O}_2$); such hosts communicate with each other within their own networks.
It is straightforward that if these two pairs of hosts exchange different information (viz., resulting in different $Comm$) then the resulting NetFlows generated in $\mathcal{O}_1$ and $\mathcal{O}_2$ will differ. Let us focus on the case where the the pairs of hosts exchange the same information (viz., same $Comm$). For simplicity, assume that the first host of each pair ($h^1$) sends \textit{exactly} the same file of 100MB to the second host, using \textit{exactly} the same protocol and ports. Let us assume that the hosts in $\mathcal{O}_1$ are allocated a bandwidth of $b_1$~Mb/s, and that those in $\mathcal{O}_2$ are allocated a bandwidth of $b_2$~Mb/s. Finally, let us assume that the organizations use the same NetFlow generation software, configured to allow the maximum duration of a NetFlow to be $d_1$ for $\mathcal{O}_1$, and $d_2$ for $\mathcal{O}_2$.
We identify four scenarios.
\begin{itemize}
    \item $b_1 \! = \! b_2$ and $d_1 \! = \! d_2$ $\rightarrow$ same $NetId$ and same $Conf$ (viz. same $Env$). For instance, if $b_1 \! = b_2 \! = \! 100$Mb/s and $d_1 \! = d_2 \! = \! 10$s, then the file will be transferred in the same amount of time (8s) in both $\mathcal{O}_1$ and $\mathcal{O}_2$, resulting in similar NetFlows (with a duration of 8s).
    \item $b_1 \! \neq \! b_2$ and $d_1 \! = \! d_2$ $\rightarrow$ different $NetId$ but same $Conf$ (viz. different $Env$). For instance, if $b_1 \! = \! 100$Mb/s and $b_2 \! = \! 1$Mb/s, then the file will be transferred in 8s in $\mathcal{O}_1$ but in 800s in $\mathcal{O}_2$, resulting in different NetFlows.
    \item $b_1 \! = \! b_2$ and $d_1 \! \neq \! d_2$ $\rightarrow$ same $NetId$ but different $Conf$ (viz. different $Env$). For instance, if $b_1 \! = b_2 \! = \! 100$Mb/s while $d_1 \! = \! 10$s and $d_2 \! = \! 1$s, then the transfer will take 8s in both $\mathcal{O}_1$ and $\mathcal{O}_2$; but in $\mathcal{O}_1$ there will be 1 NetFlow of 8s, while in $\mathcal{O}_2$ there will be 8 NetFlows of 1s.
    \item $b_1 \! \neq \! b_2$ and $d_1 \! \neq \! d_2$ $\rightarrow$ different $NetId$ and different $Conf$ (viz. different $Env$). This is self-explanatory.
\end{itemize}
Of course, there are many other factors that affect $NetId$ and $Conf$ (aside from the bandwidth and maximum duration). The above-mentioned example is just for demonstrative purposes.

\section{Guidelines for Standardize}
\label{app:recommendations}
To avoid generating network specific artifacts (cf. §\ref{ssec:standardize}), we provide some recommendations on three common NetFlow fields: the \textit{IP addresses}, the \textit{service ports}, and the \textit{duration}.

\textbf{IP addresses.}
There are two issues that may arise when standardizing the IP addresses of two distinct datasets:
\begin{itemize}
    \item different networks use different subnet masks. For instance, the internal IP addresses of $D_i$ may present the structure ``192.168.x.x'', whereas those in $D_j$ are ``175.32.x.x'';
    \item the malicious traffic of a given dataset may be entirely produced by just few machines. 
\end{itemize}
Neglecting these issues may result in ML models that distinguish legitimate from anomalous samples on the sole basis of the IP address of a host, without giving the due importance to the remaining traffic characteristic. This is a problem because if a \textit{real} attack involves a machine with a different IP address, the detector would never identify it.
We hence propose to standardize each dataset by separating \textit{internal} from \textit{external} hosts. The ML model will use these features, instead of the IP addresses, to perform its analyses. The information to perform this separation can be obtained either from the documentation of a dataset, or by inferring it from the data using expert knowledge; if such information is not obtainable, then we suggest not to use any IP-related feature.

\textbf{Service Ports.}
Handling the service ports of distinct datasets presents similar issues to the IP addresses discussed above: different networks may adopt different port policies; and the attacks captured by a given dataset may rely just on a restricted (or unique) set of ports. 
We thus propose to standardize each dataset by categorizing each port on the basis of the IANA guidelines, i.e., \textit{well-known} [0-1023], \textit{registered} [1024-49151] and \textit{dynamic} [49151-65535]. 

\textbf{Duration.}
Besides verifying that all datasets use the same measurement units, standardizing the NetFlow duration ($d$) of distinct datasets is a challenging task. On the one hand, datasets may be created with different NetFlow tools and/or different configuration parameters. For example, setting the maximum duration ($d_{max}$) of a NetFlow to 1000 or 100 seconds would lead to significantly different results\footnote{This issue can be overcome if the datasets are provided in PCAP format by properly setting the NetFlow generation tool.}. On the other hand, there may be some underlying traits of a given network that lead its machines to generate flows of different duration. To address these issues, we propose three possible solutions, all involving the identification of the smallest maximum duration across all datasets, $min(d_{max})$:
\begin{itemize}
    \item \textit{Outlier removal}. This approach assumes that (i)~the duration of the majority of samples (from all considered datasets) falls within a reduced range $[d_l, d_t]$, and that (ii)~the top limit $d_t$ of this range is \textit{lower} than $min(d_{max})$. In these circumstances, it is possible to remove the few ``outliers'' that have extremely high durations with respect to the remaining samples. 
    Despite the consequential loss of samples, removing outliers does not necessarily reduce the prediction performance. 
    \item \textit{Threshold setting}. This solution avoids data loss problems. If a dataset $D_i$ has $d_{max} \! \gg \! min(d_{max})$, its samples having $d\!>\!min(d_{max})$ will have their duration set to $min(d_{max})$. However, it is important to store the original value of $d$ if it is needed to compute some derived metrics, such as the \textit{packets per second}. This approach may be unpractical for ML leveraging sequential analyses as it disrupts the sequence of samples. 
    \item \textit{Flow splitting}. This technique enables the application of sequential ML methods. The intuition is to \textit{split} those flows that exceed $min(d_{max})$. Given a $D_i$ with $d_{max}\!>\! min(d_{max})$, the idea is to truncate all flows of $D_i$ with duration $d\!>\!min(d_{max})$ into multiple flows. As a practical example assuming duration expressed in seconds, if $min(d_{max})$=300 and $D_i$ has $d_{max}$=1000, and if a given flow in $D_i$ has $d$=700, the approach truncates this flow in three distinct flows, with $d$=(300, 300, 100). When performing the split, it is important to also update some metrics, such as the transferred bytes or packets (which can be adjusted proportionally) as well as the start and finishing times of the flow.
\end{itemize}
We observe that, in our experiments, we adopt the \textit{outlier removal} strategy. Despite being lossy, such technique still allows to devise ML-NIDS with performance matching the state-of-the-art (see Table~\ref{tab:baseline} and compare it with Table~\ref{tab:dataset}).

\input{sections/app_symbol}

%% file: sections/app_symbol.tex
\section{Symbol Table}
\label{app:symbol}

To facilitate the readability, we report in Table~\ref{tab:symbol} the major notation used throughout the main sections of our paper.

We also further explain the difference between some of our symbols introduced in §\ref{sec:idea}, and specifically the difference between the arrays and sets (e.g., \smallarray{t} and \smallset{t}). 
Let us assume a scenario where $n$=3 and $\mu$=3, meaning that $\mathbb{M}$ is a 3x3 matrix. We use the \textit{ordered arrays} \smallarray{t}, \smallarray{\tau} (or \smallarray{e}, \smallarray{\varepsilon}) to answer the question ``which elements of $\mathbb{M}$ are included in $T$ (or $E$)?''. 
A possibility is that \smallarray{t}=(1,1,2) and that \smallarray{\tau}=(2,3,3). This means that $T$ will contain $M_1^2$, $M_1^3$, $M_2^3$. Hence, \smallset{t}=(1,2) because \smallset{t} is the \textit{set} denoting the (unique) `malicious' networks included in $T$. At the same time, \smallset{\tau}=(2,3) because \smallset{\tau} is the \textit{set} denoting the (unique) attacks included in $T$. 

Finally, we stress that, in our cross-evaluation model, \smallset{o}=\smallarray{o}=$o$, because the origin of the benign samples must be the same for both the training and evaluation partitions (i.e., $T$ and $E$, respectively).

\begin{table}[!htbp]
    \centering
    \caption{Table of relevant notation used in this paper.}
    \label{tab:symbol}
    \resizebox{0.99\columnwidth}{!}{
        \begin{tabular}{c|l|c}
             \toprule
             \textsc{Symbol} &~~\textsc{Description} & \textsc{Ref}\\
             \toprule
             
             $\mathcal{O}$ & An \textit{organization} & §\ref{sssec:usecase} \\
             $o$ & The \textit{network} of the organization $\mathcal{O}$ & §\ref{sssec:usecase} \\ 
             $N$ & A set of \textit{benign} network samples & §\ref{sssec:usecase} \\ 
             $M$ & A set of \textit{malicious} network samples & §\ref{sssec:usecase} \\ 
             \midrule 
             
             $\mathbb{D}$ & A \textit{collection} of datasets, each generated in a specific network & §\ref{ssec:principles} \\
             $n$ & The \textit{cardinality} of $\mathbb{D}$, i.e., the number of different networks included in $\mathbb{D}$ & §\ref{ssec:principles} \\
             $D_i$ & The set of samples generated by network $i$ ($\bigcup_i \! D_i \! = \! \mathbb{D}$) & §\ref{ssec:principles} \\
             $N_i$, $M_i$ & The set of benign and malicious samples included in $D_i$ ($N_i\! \cup \! M_i \! = \! D_i)$ & §\ref{ssec:principles} \\
             $\mu$ & The number of all attacks included in $\mathbb{D}$ & §\ref{ssec:principles} \\
             $M_i^\alpha$ & The samples of network $i$ corresponding to the attack $\alpha$ ($\bigcup_\alpha \! M_i^\alpha \! = \! M_i$) & §\ref{ssec:principles} \\
             $\mathbb{N}$, $\mathbb{M}$ & The collection of all $N$ and $M$ included in $\mathbb{D}$ & §\ref{ssec:principles} \\
             
             \midrule
             
             $T$, $E$ & The \textit{training} and \textit{evaluation} sets of a ML-NIDS & §\ref{ssec:benefits} \\
             $N_o$ & The benign samples (from the same network $o$) used in both $T$ and $E$ & §\ref{ssec:benefits} \\
             $M_t^\tau$ & An element of $\mathbb{M}$ used in $T$ & §\ref{ssec:benefits} \\
             $M_e^\varepsilon$ & An element of $\mathbb{M}$ used in $E$ & §\ref{ssec:benefits} \\
             \scriptset{t}, \scriptset{e} & The set of all networks included in $T$ and $E$ & §\ref{ssec:benefits} \\
             \scriptset{\tau}, \scriptset{\varepsilon} & The set of all attacks included in $T$ and $E$ & §\ref{ssec:benefits} \\
             $T$(\scriptset{o}, \scriptset{t}, \scriptset{\tau}) & The function describing the training set $T$ & §\ref{ssec:benefits} \\
             $E$(\scriptset{o}, \scriptset{e}, \scriptset{\varepsilon}) & The function describing the evaluation set $E$ & §\ref{ssec:benefits} \\
             \captionC{}(\scriptset{o}, \scriptset{t}, \scriptset{e}, \scriptset{\tau}, \scriptset{\varepsilon}) & A \textit{context} is defined by the relationships between \scriptset{o}, \scriptset{t}, \scriptset{\tau}, \scriptset{e}, \scriptset{\varepsilon} & §\ref{ssec:benefits}\\
             
             \midrule
             
             $Comm$ & The contribution to a NetFlow of the \textit{communications} of the involved hosts & §\ref{ssec:standardize}\\ 
             $Env$ & The contribution to a NetFlow of the network \textit{environment} of its two hosts & §\ref{ssec:standardize}\\ 
             $NetId$ & The \textit{intrinsic properties} of a network influencing $Env$ & §\ref{ssec:standardize}\\ 
             $Conf$ & The \textit{configuration} of the NetFlow appliance influencing $Env$ & §\ref{ssec:standardize}\\ 
             $\mathbb{T}$, $\mathbb{E}$ & The set of all $T$ and $E$ generated by XeNIDS & §\ref{ssec:contextualize}\\

             \bottomrule

        \end{tabular}
    }
\end{table}